\documentclass[aps,prd,twocolumn,nofootinbib,superscriptaddress]{revtex4-1}

\usepackage{amsmath,amssymb}
\usepackage{graphicx}
\usepackage{hyperref}
\hypersetup{colorlinks=true,allcolors=blue}

\begin{document}

\title{Numerical tests of chiral fermion weights with $\eta$-invariant phases in the 3450 model}

\author{Tatsuhiro Misumi}
\email{misumi@phys.kindai.ac.jp}
\affiliation{Department of Physics, Kindai University, Higashi-Osaka, Osaka 577-8502, Japan}

\begin{abstract}
We test numerically a recently proposed prescription for coupling continuum chiral fermions to interpolated lattice gauge fields, in which the fermion weight has a Pauli--Villars (PV) regularized magnitude and a phase given by an Atiyah--Patodi--Singer (APS) $\eta$-invariant in one higher dimension. We use the anomaly-free two-dimensional $U(1)$ theory with left-handed charges $3,4$ and right-handed charges $5,0$, whose torus determinant is known exactly. At zero magnetic flux, with gauge fields interpolated from lattice links, the Polyakov loops evaluated on a $16\times16$ grid of flat connections agree with the exact values within $0.04\%$, whereas omitting the APS phase substantially reduces them to about $3\%$ of the exact result. The vacuum-polarization response yields an induced mass $m/e=2.82089$, close to $5/\sqrt\pi=2.82095$. At unit flux, where zero modes make the determinant vanish, we calculate a twelve-point correlator using the generalized APS boundary condition on a cylinder. For three smooth backgrounds its magnitude agrees with the exact result within $0.8\%$. Including the $\eta$-invariant phase reduces the correlator phase error from $1.8$--$3.0$ mrad to below $10^{-6}$ rad. For a gauge field reconstructed from lattice links, the APS phase agrees with the required value within $0.12\%$ at three different Landau cutoffs with a common heat regulator. The phase of each charged fermion changes by $17$--$48$ mrad under a change of the bulk extension, while the change cancels to below $0.02$ mrad in the anomaly-free combination.
\end{abstract}

\maketitle

\section{Introduction}
\label{sec:intro}

Defining a chiral gauge theory nonperturbatively is a longstanding problem. Fermion doubling makes it difficult to construct a chiral spectrum on a local lattice~\cite{Nielsen:1980rz}. Domain-wall and overlap fermions, together with lattice chiral symmetry, have provided important ways to address this difficulty \cite{Kaplan:1992bt,Shamir:1993zy,Narayanan:1993sk,Narayanan:1994gw,Ginsparg:1981bj,Neuberger:1997fp,Luscher:1998pqa}. In a chiral gauge theory, however, one must also define the fermion measure consistently with gauge invariance.

Gauge-invariant fermion measures have been constructed for anomaly-free Abelian theories and studied in perturbative and more general non-Abelian settings \cite{Luscher:1998du,Luscher:1999un,Luscher:2000zd,Kikukawa:2001mw,Kadoh:2003ii,Kadoh:2004uu,Kadoh:2007wz,Kadoh:2007xb,Craig:2026soq}. Other approaches use mirror-fermion decoupling, gauge-field flow, or additional dimensions \cite{Poppitz:2007tu,Giedt:2007qg,Poppitz:2010at,Chen:2012di,Grabowska:2015qpk,Grabowska:2016bis,Fukaya:2016ofi,Grabowska:2026ham}. Recent studies have also explored the construction of chiral gauge theories on manifolds with a single connected boundary~\cite{Kaplan:2023pxd,Kaplan:2023pvd,Kaplan:2024ezz,Clancy:2024bjb,Sen:2026pnm}. The two-dimensional $3450$ theory has also been studied using Hamiltonian and other nonperturbative methods \cite{Berkowitz:2023pnz,Morikawa:2024zyd,Onogi:2025tev,Thorngren:2026ydw,Lu:2026itw,Seifnashri:2026sol,Lootens:2026sim,Zeng:2022grc,Chi:2026jqr}.

Another approach keeps the gauge links on a lattice but couples them to continuum fermions through a smooth interpolation of the gauge field~\cite{Flume:1981cw,Gockeler:1992ig,Gockeler:1992bi,tHooft:1994jaf,Hsu:1995qm,Kronfeld:1995ws,Hernandez:1995jc,Bodwin:1995rd,Hernandez:1996rd}. Earlier numerical studies of two-dimensional chiral models used such interpolated fields and evaluated determinant magnitudes with Pauli--Villars (PV) regularization and Fourier cutoffs \cite{Montvay:1995qb,Gattringer:1995du,Gattringer:1996ke}. Here we also calculate the complex phase, so that the full fermion weight can be tested against the exact result.

Recent proposals give a prescription for continuum chiral fermions coupled to interpolated lattice gauge fields~\cite{Hsu:2026gmi,Witten:2026ezr}. In the prescription tested here, the magnitude of the Weyl determinant is defined using PV regularization, while its phase is determined by an Atiyah--Patodi--Singer (APS) eta invariant in one higher dimension~\cite{Atiyah:1975jf,Dai:1994kq,Witten:2019bou}. The connection between chiral determinant phases and spectral asymmetry has also been studied in related continuum and domain-wall constructions \cite{Alvarez-Gaume:1984zst,Kaplan:1995pe,Tatsumi:1999jlf,Bornyakov:1999be}. Our numerical test evaluates the PV magnitude and APS phase separately and compares their product with the exact answer.

We test this prescription in the anomaly-free two-dimensional $U(1)$ theory with left-handed charges $3,4$ and right-handed charges $5,0$, known as the $3450$ model. Exact continuum results are available for comparison~\cite{Narayanan:1996cu,Klinkhamer:2000id}. At zero magnetic flux, we construct complex fermion weights for gauge fields interpolated from lattice links. The resulting Polyakov loops agree with the exact values, whereas omitting the APS phase gives a markedly different result. We also reproduce the induced gauge-field mass. At one unit of flux, fermion zero modes make the determinant vanish. We instead calculate a twelve-point correlator using the generalized APS boundary condition. For three smooth backgrounds its magnitude agrees with the exact result within $0.8\%$, and the eta invariant reproduces the phase correction required by the choice of zero-mode frames within $0.03\%$. We also find the expected cancellation of bulk-extension dependence in the anomaly-free combination. For a $Q=1$ gauge field reconstructed from lattice links, the APS phase agrees with the required value at the subpercent level.

This paper is organized as follows. Section~\ref{sec:setup} defines the model, the gauge interpolation, and the two spectral calculations. Sections~\ref{sec:toron} and \ref{sec:nonflat} present the zero-flux determinant tests, and Sec.~\ref{sec:mass} gives the induced mass. Section~\ref{sec:qone} then tests the twelve-point correlator at one unit of flux. We discuss the results and limitations in Sec.~\ref{sec:discussion}. The appendices contain the numerical and interpolation details.

\section{Model and method}
\label{sec:setup}

\subsection{Fields and magnetic flux}

We work on a Euclidean torus $T^2$ with coordinates $0\le x,y<1$. The gauge links live on a square lattice, but the fermions remain continuum fields. Each link specifies the holonomy of the continuum gauge field along a lattice edge. We describe the interpolation between edges below. The two left-handed Weyl fermions have charges $q=3,4$, while the right-handed fermions have charges $q=5,0$. We impose antiperiodic fermion boundary conditions along both cycles, with the appropriate gauge-bundle transition conditions when the magnetic flux is nonzero. The perturbative gauge anomaly cancels because
\begin{equation}
3^2+4^2-5^2-0^2=0.
\label{eq:anomaly}
\end{equation}
There are equally many left- and right-handed fields, so the gravitational anomaly also cancels.  The neutral right-handed fermion has no gauge-field dependence and drops out of normalized gauge observables.  It is retained in the definition of the theory.

The gauge variables are $U_\mu(i,j)\in U(1)$, with $\mu=x,y$ and lattice sites $(i,j)\in\mathbb Z_{N_x}\times\mathbb Z_{N_y}$. We write $U_\mu(i,j)=\exp(i\theta^\mu_{ij})$, where $\theta^\mu_{ij}$ is a real representative of the link angle. The positively oriented plaquette based at $(i,j)$ is
\begin{align}
U_{ij}={}&U_x(i,j)U_y(i+1,j)\nonumber\\
&\times U_x(i,j+1)^{-1}U_y(i,j)^{-1}
=e^{i\phi_{ij}},
\label{eq:plaquette_definition}
\end{align}
with its principal plaquette angle defined by
\begin{equation}
\phi_{ij}=\operatorname{Arg}U_{ij}\in(-\pi,\pi].
\end{equation}
Thus $\phi_{ij}$ measures the gauge flux through cell $(i,j)$. The link angles satisfy
\begin{equation}
\theta^x_{ij}+\theta^y_{i+1,j}
-\theta^x_{i,j+1}-\theta^y_{ij}
=\phi_{ij}\pmod{2\pi}.
\label{eq:plaquette_lifts}
\end{equation}
We use the Wilson gauge action
\begin{equation}
S_g[U]=\beta\sum_{i,j}(1-\cos\phi_{ij}).
\label{eq:wilson}
\end{equation}
The parameter $\beta$ fixes the gauge coupling, but it does not enter the fermion determinant at a specified background field. The lattice magnetic flux~\cite{Flume:1981cw} is
\begin{equation}
Q=\frac{1}{2\pi}\sum_{i,j}\phi_{ij}\in\mathbb Z.
\label{eq:flux}
\end{equation}
We study two sectors, $Q=0$ and $Q=1$. Most of the flat-connection, nonflat-determinant, and polarization results use $3\times3$ links in the $Q=0$ sector. We also consider selected $Q=0$ backgrounds on $4\times4$ and $6\times6$ lattices. At $Q=1$, we consider both smooth continuum gauge fields and their link-interpolated counterparts on a $6\times6$ lattice. The principal-value choice for $\phi_{ij}$ fixes the angle branch. No admissibility bound~\cite{Luscher:1998du} is imposed on the link variables.

For comparison with the continuum result, we use the usual continuum normalization of the Wilson gauge action for smooth gauge fields,
\begin{equation}
S_g\longrightarrow\frac{1}{2e^2}\int_{T^2}d^2x\,F_{12}^2.
\label{eq:gauge_cont}
\end{equation}
For an $N\times N$ lattice of unit physical length, the spacing is $a=1/N$. The quadratic expansion of Eq.~(\ref{eq:wilson}) gives $e=N/\sqrt{\beta}$, and hence $e=3/\sqrt{\beta}$ for the $3\times3$ polarization backgrounds. Our induced-mass calculation determines the dimensionless ratio $m/e$ and is independent of the choice of $\beta$.

\subsection{Gauge interpolation}
\label{subsec:gauge_interp}

To apply a continuum Dirac operator, we must first turn the lattice links into a smooth gauge connection~\cite{Gockeler:1992bi,tHooft:1994jaf,Hernandez:1996rd}. We begin with $Q=0$, where a globally periodic gauge potential can describe the field. For the chosen links, we select real angles that satisfy Eq.~(\ref{eq:plaquette_lifts}) as an equality and interpolate each plaquette separately.

Let $s,t\in[0,1]$ be coordinates inside plaquette $(i,j)$.  We choose a smooth increasing function $S$ with $S(0)=0$ and $S(1)=1$, and write $w=S'$.  The function $w$ vanishes in a small region near each end of the unit interval and obeys
\begin{equation}
\int_0^1ds\,w(s)=1.
\label{eq:bump_norm}
\end{equation}
With suitable real lifts $\theta^x_{ij},\theta^y_{ij}$ of the link angles, the connection inside the cell is
\begin{align}
A_s&=w(s)\bigl[(1-S(t))\theta^x_{ij}
                  +S(t)\theta^x_{i,j+1}\bigr],\nonumber\\
A_t&=w(t)\bigl[(1-S(s))\theta^y_{ij}
                  +S(s)\theta^y_{i+1,j}\bigr].
\label{eq:interpolation}
\end{align}
The resulting field strength is particularly simple,
\begin{equation}
F_{st}(s,t)=\phi_{ij}w(s)w(t).
\label{eq:cell_curvature}
\end{equation}
In physical coordinates $x=(i+s)/N_x$ and $y=(j+t)/N_y$, the curvature is $F_{12}(x,y)=N_xN_yF_{st}(s,t)$. The construction is given in more detail in Appendix~\ref{app:interp}. It gives a smooth connection across cell boundaries and has the prescribed link holonomies.  A lattice gauge transformation changes the interpolated connection by a smooth continuum gauge transformation when compatible real lifts of the link and vertex angles are used.

The link angles therefore determine a continuum curvature, not just a set of plaquette averages. This distinction matters because the fermions respond to the interpolated field between lattice sites. Define the Fourier transform of the cell profile by
\begin{equation}
\widehat w(\nu)=\int_0^1ds\,w(s)e^{-2\pi i\nu s}.
\end{equation}
For integer $k=(k_x,k_y)$, the Fourier coefficient of the physical curvature is
\begin{align}
F_k={}&\widehat w(k_x/N_x)\widehat w(k_y/N_y)\nonumber\\
&\times\sum_{i,j}\phi_{ij}
e^{-2\pi i(k_xi/N_x+k_yj/N_y)}.
\label{eq:Fk}
\end{align}
In the $Q=0$ sector $F_0=0$. We reconstruct the transverse field as $A_\mu^T=\epsilon_{\mu\nu}\partial_\nu\varphi$, with $\epsilon_{xy}=+1$ and zero-average $\varphi$. For $k\ne0$ its Fourier coefficients are $\varphi_k=F_k/(4\pi^2|k|^2)$. After subtracting the transverse field, the holonomies around the two torus cycles fix the constant components of the connection, which we write as $2\pi h_\mu$. The variables $h_x,h_y$ are called \emph{torons} and are defined modulo integers. The remaining gradient is a gauge choice. We set it to zero unless we explicitly test a gauge transformation. The steps from links to a continuum operator are as follows. We construct $F_k$, recover the transverse gauge field, and retain a finite set of Fourier modes in the Dirac matrix.

We choose a smoothing width of $0.05$ in each cell. Its explicit smooth step function is given in Appendix~\ref{app:interp}. The interpolation affects the continuum curvature at fixed lattice spacing, so the exact and numerical calculations must use the same choice. The numerical operator also retains only finitely many Fourier modes of this background. At $Q=1$, we add globally defined fluctuations to a fixed reference connection carrying one unit of flux. The link interpolation in this sector is described in Sec.~\ref{subsec:qone_setup}. This preserves the nontrivial gauge bundle. A globally periodic gauge potential cannot represent $Q=1$.

\subsection{PV magnitude and APS phase}

Let $A[U]$ denote the continuum field obtained by interpolation.  In the zero-flux sector, the fermion determinant is generally nonzero. We calculate its magnitude and phase separately and combine them into the complex weight~\cite{Hsu:2026gmi,Witten:2026ezr}
\begin{align}
W[U]&=W_{\rm PV}[U]\exp\bigl[-i\pi\eta_Y(A[U])\bigr],\nonumber\\
W_{\rm PV}[U]&=\prod_{q=3,4,5}
 \bigl|\det\mathcal D_{L,q}[A[U]]\bigr|_{\rm PV}.
\label{eq:weight}
\end{align}
Here $W_{\rm PV}$ is the positive determinant magnitude of the charged fermions in the $Q=0$ sector. We use covariant derivatives $\nabla_\mu=\partial_\mu+iqA_\mu$. For our left-handed convention the Weyl operator is
\begin{equation}
\mathcal D_{L,q}=\nabla_x-i\nabla_y.
\label{eq:weyl_operator}
\end{equation}
For $q>0$ and positive magnetic flux, its index is $\operatorname{ind}\mathcal D_{L,q}=qQ$. The opposite-chirality operator is $\mathcal D_{R,q}=\nabla_x+i\nabla_y =-\mathcal D_{L,q}^{\dagger}$, with index $-qQ$. The right-handed determinant has the complex-conjugate phase, but the same positive singular-value contribution in the zero-flux sector. To obtain the magnitude, we find the eigenvalues $\lambda_n\ge0$ of $\mathcal D_{L,q}^\dagger\mathcal D_{L,q}$ for each charge $q$. At finite spectral cutoff and PV mass $M$, the contribution is
\begin{equation}
\log W_{{\rm PV},q}
=\frac12\sum_n^{\rm cutoff}
 \log\frac{\lambda_n}{\lambda_n+M^2}
 +C_q,
\label{eq:pv_product}
\end{equation}
where the constant $C_q$ does not depend on the background and cancels in the ratios studied here. For $Q=0$ these operators have no zero modes at the generic backgrounds studied below. At $Q=1$ the exact zero eigenvalues are omitted from Eq.~(\ref{eq:pv_product}) and handled explicitly through fermion insertions.

The magnitude contains no information about the complex phase. We obtain that phase from the spectral asymmetry of a Hermitian three-dimensional Dirac operator $H_Y$ on an auxiliary manifold $Y$~\cite{Alvarez-Gaume:1984zst,Witten:2019bou}. For $Q=0$, we use the solid torus $Y=D^2\times S^1_y$. The $x$ cycle bounds the disk, while the $y$ cycle remains a circle. We choose a smooth metric with product form near $\partial Y=T^2$ and extend the boundary gauge field smoothly into the disk. For the backgrounds with an invertible boundary Weyl operator considered here, we diagonalize the tangential Dirac operator entering the APS boundary condition and remove its negative-eigenvalue subspace with the APS spectral projector. This gives the boundary condition for the Hermitian three-dimensional spectral problem. The geometry, bulk connection, and the radial basis are specified in Appendix~\ref{app:aps}. For a fixed charge $q$, let $\xi_n$ be the real eigenvalues of $H_Y$. We define the single-charge eta invariant by
\begin{align}
\eta_q(s)&=\sum_{\xi_n\ne0}
  \operatorname{sgn}(\xi_n)|\xi_n|^{-s},\nonumber\\
\eta_q&=\eta_q(0)+\dim\ker H_Y,
\label{eq:eta_definition}
\end{align}
where the value at $s=0$ is obtained by analytic continuation. The kernel term follows Witten's convention. The spectral condition follows the APS construction \cite{Atiyah:1975jf,Dai:1994kq,Witten:2026ezr}. The eta invariant in Eq.~(\ref{eq:weight}) is the signed sum $\eta_Y=\eta_3+\eta_4-\eta_5-\eta_0$, with opposite signs for opposite chiralities. The neutral contribution cancels from normalized gauge-field ratios. The generic backgrounds studied here have no exact bulk zero modes. The orientation and zero-eigenvalue conventions are given in Appendix~\ref{app:aps}.

The two factors in Eq.~(\ref{eq:weight}) are evaluated separately. For the determinant magnitude at $Q=0$ we use a Fourier--Galerkin representation with antiperiodic momenta $2\pi(n_\mu+1/2)$, $|n_\mu|\le K$, and PV mass $M$~\cite{Montvay:1995qb,Gattringer:1995du,Gattringer:1996ke}. We work with $\widetilde D_q=\mathcal D_{R,q}=-\mathcal D_{L,q}^{\dagger}$, whose square Fourier matrix has the same singular values as that of $\mathcal D_{L,q}$. For given torons $h_x,h_y$, it is the sum of the diagonal flat-field operator
\begin{equation}
(\widetilde D_{0,q})_{nn}=2\pi i(n_x+1/2+qh_x)
 -2\pi(n_y+1/2+qh_y),
\label{eq:D0}
\end{equation}
and a matrix built from the Fourier components of the nonconstant gauge field. The smooth gauge field is represented by Fourier modes $|k_x|,|k_y|\le k_{\rm cut}$. This third cutoff controls the representation of the interpolated background. For flat fields the Fourier eigenvalues are summed directly. For nonflat fields we compute the singular values of the finite matrix, and the quadratic response to the background is evaluated from matrix traces. When extrapolating the PV magnitude, we first increase $K$ at fixed $M$ and then study the large-$M$ behavior. At fixed $K$, taking $M$ too large can give substantial errors.

For the phase we expand the solid-torus spinors in angular Fourier modes and radial functions, impose the APS projector, and diagonalize the positive quadratic form $\langle H_Y\Psi,H_Y\Psi\rangle$ on the constrained space. For nonflat fields we use an angular Fourier cutoff $L$, radial basis size $p$, and heat-kernel regulator $\delta$. For flat fields the Fourier cutoff along the boundary cycles is denoted by $K_{\rm APS}$. If $\mu_j$ and $v_j$ are eigenvalues and orthonormal eigenvectors of the quadratic form, we evaluate
\begin{equation}
\eta_\delta\simeq\sum_{\mu_j>0}
\frac{\langle v_j,H_Yv_j\rangle}{\sqrt{\mu_j}}
\operatorname{erfc}(\sqrt{\delta\mu_j}).
\label{eq:eta_estimator_main}
\end{equation}
The first-order matrix elements retain the sign information missing from the spectrum of $H_Y^2$. The regulator $\delta$ provides a heat-kernel prescription for the spectral asymmetry. The spectral units used to define $\delta$ in the two flux sectors are specified in Appendices~\ref{app:aps} and \ref{app:qone}. These cutoffs are independent of the gauge-field lattice size. A relative eta is obtained by subtracting target and reference values computed at common spectral parameters. The finite-basis implementation and cutoff tests are given in Appendix~\ref{app:aps}.

For later use, a general $Q=0$ field can be decomposed as
\begin{equation}
A_\mu=2\pi h_\mu+\epsilon_{\mu\nu}\partial_\nu\varphi
                   +\partial_\mu\chi,
\label{eq:hodge}
\end{equation}
where $h_x,h_y$ are the toron variables (the constant gauge-field components, defined modulo integers), $\epsilon_{xy}=+1$, $\varphi$ has zero average, and $\chi$ is a gauge function.  The term $A_\mu^T=\epsilon_{\mu\nu}\partial_\nu\varphi$ is transverse and has no constant part. In particular, $\int_{T^2}d^2x\,A_\mu=2\pi h_\mu$ for periodic $\chi$. We use the Coulomb representative $\chi=0$ in the determinant calculation.  Gauge transformations are tested separately.

The analytic determinant in this representative factorizes into a flat-toron part and a nonflat part~\cite{Narayanan:1996cu,Klinkhamer:2000id},
\begin{equation}
\frac{W_{\rm exact}[h,\varphi]}{W_{\rm exact}[h,0]}
=\exp\left[-\frac{25}{2\pi}\int_{T^2}d^2x\,A_\mu^T A_\mu^T\right].
\label{eq:analytic_ratio}
\end{equation}
In particular, the right-hand side is real and positive.  Thus the relative phase of the combined $3450$ determinant is zero when the toron is held fixed.  Equation~(\ref{eq:analytic_ratio}), like the other exact formulas quoted below, serves only as a benchmark. No analytic determinant or correlator enters the numerical PV, zero-mode, or APS calculations.

\subsection{Zero modes at $Q=1$}
\label{subsec:qone_setup}

At $Q=1$, a massless fermion determinant without insertions vanishes, since $\operatorname{ind}\mathcal D_{L,q}=q$ by the index theorem~\cite{Atiyah:1968mp,Atiyah:1975jf}. With our chirality conventions, there are three zero modes of the left-handed charge-$3$ fermion, four of charge $4$, and five conjugate zero modes associated with the right-handed charge-$5$ fermion. An appropriate nonzero fermion integral therefore has twelve insertions,
\begin{equation}
{\cal O}_{12}(X)=
 \prod_{a=1}^3\psi_3(x_a)
 \prod_{b=1}^4\psi_4(y_b)
 \prod_{c=1}^5\bar\psi_5(z_c).
\label{eq:qone_operator}
\end{equation}
Here $X=\{x_a,y_b,z_c\}$ denotes twelve points on $T^2$. The insertions saturate the twelve zero-mode Grassmann integrals. There are no insertions of the neutral fermion. 
A similar insertion, compensating an anomaly-induced mismatch of fermion zero modes, was used in lattice simulations of two-dimensional supersymmetric Yang--Mills theory on curved backgrounds~\cite{Kamata:2016xmu}.
Writing $S_f[A]$ for the Euclidean action of the four free Weyl fields coupled to the fixed background $A$, we calculate the \emph{unnormalized} fermion integral
\begin{equation}
\mathcal C_1[A;X]=\int D\psi\,D\bar\psi\,
{\cal O}_{12}(X)e^{-S_f[A]}.
\label{eq:qone_fermion_integral}
\end{equation}
We do not divide by the $Q=1$ partition function, which is zero. The operator in Eq.~(\ref{eq:qone_operator}) is gauge covariant. For separated insertion points its individual factors transform locally. We compare it in a fixed gauge. Wilson-line transporters would be needed to form an explicitly gauge-invariant separated-point observable.

As the reference we take a unit-area torus with uniform curvature $F_{12}^{\rm ref}=2\pi$. In a local gauge, $A_x^{\rm ref}=0$ and $A_y^{\rm ref}=2\pi x$, with the usual transition function across the $x$ cycle. For a charge-$q$ spinor, our antiperiodic boundary conditions in this gauge read $\psi(x+1,y)=-e^{-2\pi iqy}\psi(x,y)$ and $\psi(x,y+1)=-\psi(x,y)$. For the backgrounds studied here, we keep the torons fixed relative to the reference and add globally periodic functions $\varphi,\chi$, with $\varphi$ having zero average,
\begin{equation}
A_\mu=A_\mu^{\rm ref}
 +\epsilon_{\mu\nu}\partial_\nu\varphi
 +\partial_\mu\chi,\qquad
F_{12}=2\pi-\partial^2\varphi.
\label{eq:qone_background}
\end{equation}
The reference and target lie in the same $Q=1$ bundle. For the $Q=1$ lattice backgrounds studied here, dividing the link variables by the corresponding holonomies of the unit-flux reference field gives links with zero principal-branch flux. We interpolate these relative links as in Sec.~\ref{subsec:gauge_interp}. Adding the interpolated field to the reference connection reproduces the original link holonomies. The spectral tests in Secs.~\ref{subsec:qone_exact}--\ref{subsec:qone_aps} use smooth continuum backgrounds with a few Fourier modes. Their link-interpolated counterparts are examined in Sec.~\ref{subsec:qone_links}. We keep the insertion points and the reference zero-mode frame fixed and evaluate the complex ratio
\begin{equation}
R_1[A;X]=
 \frac{\mathcal C_1[A;X]}
 {\mathcal C_1[A^{\rm ref};X]}.
\label{eq:qone_ratio_definition}
\end{equation}
This ratio removes the phase of the reference correlator, and with it the relative theta angle between $Q=0$ and $Q=1$, but retains the dependence on the gauge field and the insertion points.

A boundary carrying one unit of flux cannot be filled by a three-manifold carrying a smooth extension of that $U(1)$ bundle. Instead, following the zero-mode prescription of Ref.~\cite{Witten:2026ezr}, we introduce a second boundary with the reference field. We use a cylinder $Y=T^2\times[0,1]$, whose two ends carry the target and reference fields with opposite boundary orientations. Writing $\tau\in[0,1]$ for the cylinder coordinate, we use $A_\tau=0$ and
\begin{align}
A_\mu(\tau,x,y)={}&A_\mu^{\rm ref}(x,y)\nonumber\\
 &+s(\tau)\bigl[A_\mu(x,y)-A_\mu^{\rm ref}(x,y)\bigr],
\label{eq:qone_bulk_interp}
\end{align}
where $s(0)=0$, $s(1)=1$, and its derivatives vanish at the ends. We take $s(\tau)=S(\tau)$, with the smooth function $S$ defined in Appendix~\ref{app:interp}, and use the flat product metric $d\tau^2+dx^2+dy^2$ on the cylinder. The three-dimensional operator has the form $H_Y=i\Gamma^\tau\partial_\tau+B(\tau)$, where $\Gamma^\tau$ is the normal Clifford matrix and $B$ is determined by the two-dimensional Weyl matrix. We expand the cylinder spinor in $p$ Legendre polynomials in $\tau$. The tangential operator entering the APS boundary condition is $\mathcal B(\tau)=-i\Gamma^\tau B(\tau)$, so $H_Y=i\Gamma^\tau(\partial_\tau+\mathcal B)$. For the cylinder spinor $\Psi$ we impose on the nonzero boundary modes
\begin{equation*}
P_+(\mathcal B(0))\Psi(0)=0,\qquad
P_-(\mathcal B(1))\Psi(1)=0,
\end{equation*}
where $P_\pm$ project onto the positive and negative eigenspaces, respectively. The opposite signs reflect the opposite outward normals at the two ends. For each charge, the positive- and negative-chirality boundary zero-mode spaces on this combined boundary have equal dimension. The generalized APS condition identifies them by a unitary map $U:V_+\to V_-$. In our convention $U$ maps the positive-chirality target zero modes to the negative-chirality reference zero modes. We use the inverse map $V\equiv U^{-1}$ in the boundary constraint
\begin{equation}
Z_1^\dagger\Psi(1)=VZ_0^\dagger\Psi(0),
\qquad V=U^{-1},
\label{eq:qone_zero_gluing}
\end{equation}
where $Z_0,Z_1$ are orthonormal frames of the endpoint zero-mode spaces. At fixed endpoint frames, $V\to V'$ multiplies the left-handed zero-mode factor by $\det(V'V^{-1})$, with the inverse change in the eta factor. The right-handed contribution is complex conjugated. In the reference-normalized correlator, we use the same $V$ for the target and reference cylinders, so the common gluing determinant cancels. We also test the compensation directly by varying $V$ while keeping the reference normalization fixed in Sec.~\ref{subsec:qone_aps}; see Appendix~\ref{app:qone} for the implementation.

The $Q=1$ calculation uses a Landau-level basis in two dimensions, which contains the $q$ exact zero modes for each positive charge. In the uniform reference field, the nonzero squared singular values are $4\pi qn$ ($n=1,2,\ldots$). At finite level number $n_{\rm max}$, we use a rectangular Weyl matrix with $q(n_{\rm max}+1)$ columns and $qn_{\rm max}$ rows. Its index is therefore exactly $q$ even before the spectral cutoff is removed. The Landau cutoffs are expressed in units of $4\pi$ for the squared singular values of the reference operator. For the PV calculation we retain Landau levels satisfying $qn\le E$, with a common Landau cutoff $E$ for all three charges,
\begin{equation}
n_{\rm max}=\left\lfloor E/q\right\rfloor.
\label{eq:qone_energy_cutoff}
\end{equation}
We omit the exact zero modes from the PV product. The zero-mode wave functions give the Slater determinants. The $x$-dependent backgrounds used in Secs.~\ref{subsec:qone_exact}--\ref{subsec:qone_aps} leave the magnetic guiding-center label unchanged, so the Weyl matrix separates into $q$ smaller blocks. In the three-dimensional phase calculation we use a separate Landau cutoff $E_\eta$, with $n_{\rm max}=\lfloor E_\eta/q\rfloor$, along with $p$ and $\delta$. At finite Landau cutoff the PV magnitude is not exactly gauge invariant. The discarded levels produce a small dependence on the pure-gauge function $\chi$, which we measure in Sec.~\ref{sec:qone}.
Table~\ref{tab:cutoff_guide} summarizes the spectral cutoffs used in the two flux sectors.

\begin{table}[t]
\begin{ruledtabular}
\begin{tabular}{lll}
Symbol & Meaning & Sector \\
\hline
$K$ & Weyl Fourier bound & $Q=0$, PV\\
$M$ & PV mass & Both\\
$k_{\rm cut}$ & Gauge Fourier bound & Both\\
$K_{\rm APS}$ & Boundary Fourier bound & $Q=0$, flat\\
$L$ & APS angular Fourier bound & $Q=0$, nonflat\\
$p$ & APS radial / cylinder basis size & Both\\
$\delta$ & APS heat parameter & Both\\
$E$ & Landau energy bound, PV & $Q=1$\\
$E_\eta$ & Landau energy bound, APS & $Q=1$\\
$n_{\rm max}$ & Charge-dependent level bound & $Q=1$
\end{tabular}
\end{ruledtabular}
\caption{Numerical cutoffs. The symbol $p$ counts radial basis functions on the solid torus and Legendre polynomials on the cylinder; these are different bases. The level bound is $\lfloor E/q\rfloor$ for PV and $\lfloor E_\eta/q\rfloor$ for APS.}
\label{tab:cutoff_guide}
\end{table}

\section{Flat connections}
\label{sec:toron}

\subsection{Complex weight}

Let us consider a test that is sensitive to the chiral phase. For a flat connection, $F_{12}=0$, the gauge field is specified by the torons $h_x,h_y$. With antiperiodic boundary conditions, the known continuum determinant~\cite{Narayanan:1996cu,Klinkhamer:2000id} is proportional to
\begin{equation}
W_{\rm flat}(h)\propto
       z_3(h)\,z_4(h)\,\overline{z_5(h)},
\label{eq:flat_exact}
\end{equation}
where the single-charge factor is
\begin{equation}
z_q(h)=e^{i\pi q^2h_xh_y-\pi(qh_y)^2}
\frac{\vartheta_3\bigl(\pi q(h_x+ih_y)\,|\,i\bigr)}
     {\vartheta_3(0\,|\,i)},
\label{eq:theta_factor}
\end{equation}
and $\vartheta_3(z|\tau)=\sum_{n\in\mathbb Z} e^{i\pi\tau n^2+2inz}$. The neutral factor is independent of $h$ and is omitted. The bilinear phases in the three charged factors cancel by Eq.~(\ref{eq:anomaly}), but the full weight still has a nontrivial phase. The conjugation in Eq.~(\ref{eq:flat_exact}) is essential. This is because the phase cannot be recovered from the determinant magnitude alone.

For each toron we calculate the magnitude from the PV-regularized Fourier spectrum and the phase from the APS spectral problem on the solid torus, and combine them into a complex weight. The APS cutoff scans are given in Appendix~\ref{app:aps} and the PV extrapolation in Appendix~\ref{app:pv}.

\subsection{Polyakov loops}

On the $3\times3$ lattice, we define the Polyakov loops around the two cycles of the torus by
\begin{equation}
P_x=\prod_{n_x=0}^{2}U_x(n_x,n_y),\qquad
P_y=\prod_{n_y=0}^{2}U_y(n_x,n_y).
\label{eq:polyakov}
\end{equation}
For a flat connection, $P_\mu=\exp(2\pi i h_\mu)$.  Its expectation value in the toron sector is obtained by integrating the complex fermion weight over the toron variables,
\begin{equation}
\langle P_\mu\rangle_{\rm flat}
=\frac{\int_{-1/2}^{1/2}d^2h\,P_\mu(h)W_{\rm flat}(h)}
       {\int_{-1/2}^{1/2}d^2h\,W_{\rm flat}(h)}.
\label{eq:polyakov_flat}
\end{equation}
The Wilson action is constant on this flat family.  This integral should not be confused with the full integral over all gauge fields. We also compute the average phase factor of the complex weight,
\begin{align}
\left\langle e^{i\theta}\right\rangle_{|W|}
&=\frac{\int d^2h\,W_{\rm flat}(h)}
        {\int d^2h\,|W_{\rm flat}(h)|},\nonumber\\
W_{\rm flat}(h)&=|W_{\rm flat}(h)|e^{i\theta(h)},
\label{eq:average_phase}
\end{align}
with both integrals over $[-1/2,1/2]^2$.

We evaluate Eq.~(\ref{eq:polyakov_flat}) on a $16\times16$ midpoint grid of toron values, with the APS setting $(p,K_{\rm APS},\delta)=(28,28,0.0125)$ at every grid point. The phases at 90 toron values are computed directly. The others follow from the reflection identities in Eq.~(\ref{eq:aps_reflections}), which hold exactly in the finite basis and which we have checked on directly computed pairs. The analytic weight is evaluated at the same grid points. The analytic Polyakov loop is $0.803950$ on this grid and $0.803951$ for the continuous toron integral. In contrast, the average phase factor and phase-quenched Polyakov loop show a larger grid dependence because $|W|$ is not smooth at its zeros. For the analytic weight they change from $0.284800$ and $0.025269$ on the $16\times16$ grid to $0.286330$ and $0.029288$ on a $256\times256$ grid. We therefore use the same $16\times16$ midpoint grid when comparing the numerical and analytic toron integrals.

\begin{figure*}[t]
\includegraphics[width=0.92\textwidth]{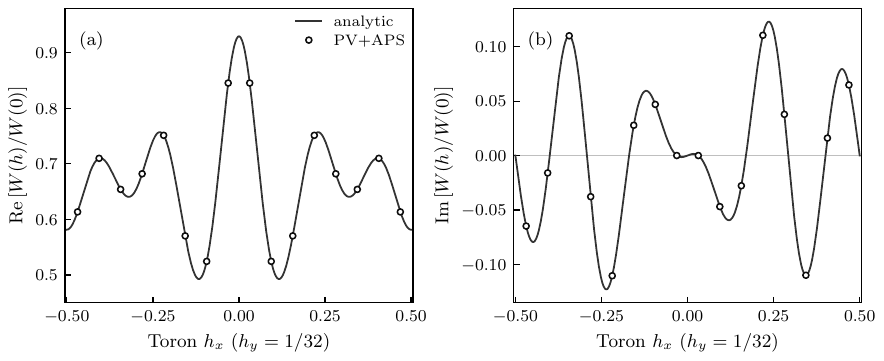}
\caption{Complex fermion weight at $h_y=1/32$. Open circles show the numerical PV+APS weight at common cutoffs; solid curves are the exact torus determinant. Both weights are normalized at $h=0$.}
\label{fig:flatweight}
\end{figure*}

Figure~\ref{fig:flatweight} compares the numerical and exact complex weights along $h_y=1/32$. The agreement extends over the full $16\times16$ grid. With both weights normalized to $W(0)=1$, the largest value of $|W_{\rm num}(h)-W_{\rm exact}(h)|$ is $1.1\times10^{-3}$, and the root-mean-square difference is $3.5\times10^{-4}$. The maximum phase difference from the analytic determinant over the 256 torons is $0.00250$ rad, and its root-mean-square value is $0.00094$ rad.

\begin{table}[t]
\begin{ruledtabular}
\begin{tabular}{lccc}
APS calculation & $\mathrm{Re}\langle P_x\rangle$ & $\mathrm{Re}\langle P_y\rangle$ & $\langle e^{i\theta}\rangle_{|W|}$\\
\hline
Lower cutoffs ($p=10$) & $0.833258$ & $0.805175$ & $0.280159$\\
Higher cutoffs ($p=28$) & $0.804221$ & $0.803787$ & $0.284776$\\
Analytic & $0.803950$ & $0.803950$ & $0.284800$
\end{tabular}
\end{ruledtabular}
\caption{Polyakov loops and average phase factor on the $16\times16$ toron grid. Each numerical row uses a single set of APS spectral cutoffs over the whole grid.}
\label{tab:polyakov}
\end{table}

Table~\ref{tab:polyakov} gives the Polyakov loops obtained with the same weights: $\mathrm{Re}\langle P_x\rangle=0.804221$ and $\mathrm{Re}\langle P_y\rangle=0.803787$ differ from the analytic grid value $0.803950$ by $0.034\%$ and $0.020\%$, respectively. At the lower setting $(p,K_{\rm APS},\delta)=(10,16,0.025)$ we find $\mathrm{Re}\langle P_x\rangle=0.833258$. The Polyakov-loop expectations are real up to tiny numerical errors on the symmetrically completed toron grid. The exact square-torus result is symmetric under interchange of the two cycles. At finite APS cutoffs this symmetry is not manifest because the solid-torus construction fills the $x$ cycle. The different errors in $P_x$ and $P_y$ are therefore numerical, not a physical anisotropy. The average phase factor is about $0.285$, showing substantial cancellation of complex weights. For the exact continuum weight, Eq.~(\ref{eq:analytic_ratio}) implies that the phase throughout the $Q=0$ sector depends only on the torons. Since the Wilson gauge action and the gauge-field measure factorize from the toron variables in this sector, the average phase factor over the full $Q=0$ gauge ensemble equals the flat-toron integral in Eq.~(\ref{eq:average_phase}). This relation has not been tested by sampling nonflat gauge fields with the finite-cutoff PV and APS weights.

If we omit the phase but keep the same determinant magnitudes, both Polyakov loops become $0.02527$ instead of $0.80395$. The determinant magnitude alone does not reproduce the observable.

\section{Nonflat connections}
\label{sec:nonflat}

For nonflat gauge fields, both the magnitude and the phase of the fermion weight can be compared with the known continuum determinant~\cite{Narayanan:1996cu,Klinkhamer:2000id}. We use the smooth interpolation of Sec.~\ref{sec:setup} and vary the plaquette angles at fixed toron.  In the Coulomb representative, Eq.~(\ref{eq:analytic_ratio}) gives the exact complex determinant ratio. Its magnitude is fixed by the transverse gauge field and its phase is zero. Using the Fourier coefficients of the interpolated field in Eq.~(\ref{eq:Fk}), the exact ratio $R_{\rm exact}=W_{\rm exact}[h,\varphi]/W_{\rm exact}[h,0]$ obeys
\begin{equation}
\log R_{\rm exact}=-\frac{25}{8\pi^3}
\sum_{k\ne0}\frac{|F_k|^2}{k_x^2+k_y^2}.
\label{eq:analytic_F}
\end{equation}
The sum includes all Fourier modes of the smooth field, unlike the finite Fourier representation used in the PV matrices.

We select two nonflat configurations on the $3\times3$ lattice, denoted by I and II. Each pattern has zero total flux and is scaled to the maximum plaquette angle shown in Table~\ref{tab:nonflat}.  The explicit plaquette arrays are given in Appendix~\ref{app:interp}. We also test one sinusoidal zero-flux background on each of the $4\times4$ and $6\times6$ lattices. Their link variables are specified in the same appendix. The numerical PV ratio is obtained directly from the continuum Dirac singular values at $K=20$, $M=24$. The APS relative phase is computed from the three-dimensional spectrum using $\delta=0.00625$ and the values of $(L,p)$ in the table.

We compare the normalized weights
\begin{align}
R(h,\varphi)&=\frac{W[h,\varphi]}{W[h,0]},\nonumber\\
R_{\rm PV}&=\frac{W_{\rm PV}[h,\varphi]}{W_{\rm PV}[h,0]},\nonumber\\
R_{\rm num}&=R_{\rm PV}\exp(i\Delta\theta_{\rm APS}).
\label{eq:nonflat_ratio}
\end{align}
Here $\Delta\theta_{\rm APS}$ is the difference between the nonflat and flat APS phases at the same toron, with the phase convention of Eq.~(\ref{eq:weight}).  The ratios remove the overall normalization of the determinant.  In the analytic answer, $R_{\rm exact}$ is real and positive, so the listed APS phase is also the relative phase error.

\begin{table*}[t]
\begin{ruledtabular}
\begin{tabular}{lccccccc}
Background & $(h_x,h_y)$ & $\max_{i,j}|\phi_{ij}|$ & $(L,p)$ & $R_{\rm exact}$ & $R_{\rm PV}$ & $\Delta\theta_{\rm APS}$ (rad) & Complex error\\
\hline
I & $(0.46875,0.03125)$ & $0.461$ & $(5,12)$ & $0.699982$ & $0.708898$ & $-0.000945$ & $1.277\%$\\
I & $(0.21875,0.28125)$ & $0.240$ & $(5,10)$ & $0.907826$ & $0.910960$ & $-0.000967$ & $0.359\%$\\
II & $(-0.15625,0.09375)$ & $0.200$ & $(7,14)$ & $0.922959$ & $0.925381$ & $+0.000038$ & $0.262\%$\\
II & $(-0.15625,0.09375)$ & $0.400$ & $(4,10)$ & $0.725653$ & $0.733301$ & $+0.015191$ & $1.855\%$\\
III ($4\times4$) & $(0.21875,0.28125)$ & $0.1125$ & $(5,10)$ & $0.864966$ & $0.868792$ & $-0.000580$ & $0.446\%$\\
III ($6\times6$) & $(0.21875,0.28125)$ & $0.0500$ & $(5,10)$ & $0.857264$ & $0.860943$ & $-0.002070$ & $0.477\%$
\end{tabular}
\end{ruledtabular}
\caption{Nonflat determinant ratios from the PV and APS calculations. The complex error is $|R_{\rm num}/R_{\rm exact}-1|$; angles and phases are in radians. All PV values use $K=20$ and $M=24$, with background Fourier cutoff $k_{\rm cut}=6$.}
\label{tab:nonflat}
\end{table*}

Table~\ref{tab:nonflat} shows that the calculated complex weights agree with the analytic determinant at the subpercent level in the two weak-curvature $3\times3$ backgrounds. The differences increase to $1$--$2\%$ for the stronger-curvature cases. The $4\times4$ and $6\times6$ backgrounds also give subpercent agreement at the same PV and APS cutoffs. At fixed $M=24$ and APS phase, increasing the fermion cutoff from $K=20$ to $24$ reduces their complex errors to $0.386\%$ and $0.420\%$, respectively.

The analytic values in Table~\ref{tab:nonflat} use the full smooth interpolation, while the numerical PV operator retains only background Fourier modes with $|k_x|,|k_y|\le6$. Truncating the analytic background in the same way changes the logarithm of its weight by less than $4.2\times10^{-4}$ in these cases. This accounts for less than $0.05\%$ of the magnitude ratio. The other differences come from the finite fermion and PV cutoffs.

\begin{figure}[t]
\includegraphics[width=0.98\columnwidth]{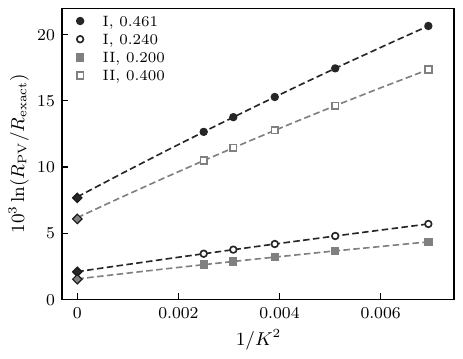}
\caption{Logarithmic deviation of the PV determinant magnitude from the analytic massless result for the four $3\times3$ nonflat backgrounds in Table~\ref{tab:nonflat}, at fixed $M=24$. Symbols are the spectral data for $K=12,14,16,18,20$; dashed curves are second-order polynomial fits in $1/K^2$. Diamonds at $1/K^2=0$ are predictions from the background truncation and the finite-$M$ response at quadratic order in the background field, Eq.~(\ref{eq:nonflat_finite_M}). Higher-order regulator effects at finite field strength are not included in the predictions.}
\label{fig:nonflat}
\end{figure}

Figure~\ref{fig:nonflat} shows the finite-$K$ dependence at $M=24$. Increasing $K$ reduces the magnitude discrepancy, but a finite-$M$ contribution remains in the limit $K\to\infty$. Second-order polynomial fits in $1/K^2$ give intercepts within about $2\%$ of the predictions from the finite-$M$ response at quadratic order in the background field, Eq.~(\ref{eq:nonflat_finite_M}). The small differences can include higher-order regulator effects as well as uncertainty in the extrapolation. At the cutoffs of Table~\ref{tab:nonflat}, this finite-$M$ term accounts for $55$--$57\%$ of the logarithmic magnitude discrepancy in the four $3\times3$ cases. The remainder contains finite-$K$ effects and higher-order terms in the massive regulator determinant, as discussed in Appendix~\ref{app:pv}.

For configuration II at $\max_{i,j}|\phi_{ij}|=0.2$, the APS relative phase decreases from $0.004054$ rad at $(L,p)=(4,8)$ to $0.0000376$ rad at $(7,14)$. The angular-cutoff dependence of configuration I is less regular. The detailed cutoff comparisons are given in Appendix~\ref{app:aps}. Since a joint limit in the angular and radial cutoffs has not been established, we report the finite cutoffs used for every complex-weight result.

We also test two properties of the APS phase at finite cutoff. Under a small smooth gauge transformation, the phase change decreases from $0.001150$ rad at $(L,p)=(3,6)$ to $0.000408$ rad at $(4,8)$. For two different bulk extensions of the same boundary field in configuration II, the phase difference is $4.80\times10^{-5}$ rad at $(7,14)$. Its cutoff dependence is not monotonic. The backgrounds, transformations, and complete comparisons are given in Appendix~\ref{app:gauge}.

\section{Induced gauge-field mass}
\label{sec:mass}

We next use the $Q=0$ PV calculation to test another physical quantity, the induced gauge-field mass. In the Schwinger model, this mass is determined by the vacuum polarization~\cite{Schwinger:1962tp}. For the $3450$ theory, each charged Weyl fermion contributes half of the corresponding Dirac vacuum polarization.  The total transverse polarization is
\begin{equation}
\Pi_{\mu\nu}(k)=\frac{25}{\pi}
  \left(\delta_{\mu\nu}-\frac{k_\mu k_\nu}{k^2}\right),
\label{eq:polarization}
\end{equation}
where the gauge potential does not include a factor of $e$. The physical Fourier momentum is $2\pi k$. Combined with Eq.~(\ref{eq:gauge_cont}), this gives
\begin{equation}
\frac{m^2}{e^2}=\frac{25}{\pi},\qquad
\frac{m}{e}=\frac{5}{\sqrt\pi}=2.8209479177\ldots.
\label{eq:mass_exact}
\end{equation}
The transverse form of Eq.~(\ref{eq:polarization}) respects gauge invariance. It is not a bare Proca mass.

We determine this coefficient numerically. To this end, we take a one-parameter family of nonflat gauge fields on the $3\times3$ lattice,
\begin{align}
U_x(n_x,n_y)&=1,\nonumber\\
U_y(n_x,n_y)&=\exp\left[it\cos\left(\frac{2\pi n_x}{3}\right)\right].
\label{eq:mass_probe}
\end{align}
The plaquette angles are $-3t/2$, $0$, and $3t/2$, each repeated three times.  The total flux and the average toron are zero.  This background has no constant transverse gauge potential, and its physical normalization is fixed by the link holonomies.

We define the positive geometric quantity
\begin{equation}
I=\int_{T^2}d^2x\,
       \left|\left.\frac{\partial A_\mu^T}{\partial t}\right|_{t=0}\right|^2
  =2.6102879182\ldots,
\label{eq:geom_I}
\end{equation}
and extract the coefficient $\mathcal Q_{\rm PV}$ from the spectral calculation
\begin{equation}
-\log\frac{W_{\rm PV}[t]}{W_{\rm PV}[0]}
 =\mathcal Q_{\rm PV}t^2+O(t^3).
\label{eq:mass_Q}
\end{equation}
The measured mass coefficient is then
\begin{equation}
c_{\rm PV}=\frac{2\mathcal Q_{\rm PV}}{I},\qquad
\left(\frac{m}{e}\right)_{\rm PV}=\sqrt{c_{\rm PV}}.
\label{eq:mass_extract}
\end{equation}
The geometric integral $I$ is determined by the Fourier modes of the interpolated curvature alone, as shown in Appendix~\ref{app:pv}. The spectral calculation retains background Fourier modes with $|k_x|,|k_y|\le k_{\rm cut}$. Let $I_{k_{\rm cut}}$ be the same transverse norm with only these modes retained. The analytic response to the truncated background is
\begin{align}
\mathcal Q_{\rm analytic}(k_{\rm cut})
 &=\frac{25}{2\pi}I_{k_{\rm cut}},\nonumber\\
\left(\frac{m}{e}\right)_{\rm analytic,\,k_{\rm cut}}
 &=\frac{5}{\sqrt\pi}\sqrt{\frac{I_{k_{\rm cut}}}{I}}.
\label{eq:mass_finite_background}
\end{align}
The second expression uses the full norm $I$ in the denominator, as does Eq.~(\ref{eq:mass_extract}). It is therefore the appropriate analytic comparison for the mass ratio extracted at finite $k_{\rm cut}$. Its difference from $5/\sqrt\pi$ is caused solely by the Fourier truncation of the background field. Appendix~\ref{app:pv} gives the finite-matrix trace formula used to calculate $\mathcal Q_{\rm PV}$.

We vary the fermion Fourier cutoff $K$, the regulator mass $M$, and the background Fourier cutoff $k_{\rm cut}$. In the PV-subtracted quadratic response, the integrand falls as $M^2/|n|^4$ at large fermion mode number $n$ and fixed $M$, giving a leading Fourier-cutoff correction of order $M^2/K^2$. The finite-$M$ polarization also has an expansion in $1/M^2$ at fixed background momentum; see Eq.~(\ref{eq:finite_M_response}). Motivated by these behaviors, we extrapolate linearly in $1/K^2$ at each $M$, followed by a quadratic fit in $1/M^2$ using the four largest masses. We vary the fit windows and cutoff ranges to assess the uncertainty of the extrapolation.

The extracted mass approaches the continuum value as additional Fourier modes of the interpolated background are retained. We first study seven background cutoffs with common fermion cutoffs $K\le220$, then extend the calculation to $K=500$ and $M=128$. The full range of cutoffs is examined in Appendix~\ref{app:pv}, where representative fitted results are tabulated.

\begin{figure*}[t]
\includegraphics[width=0.93\textwidth]{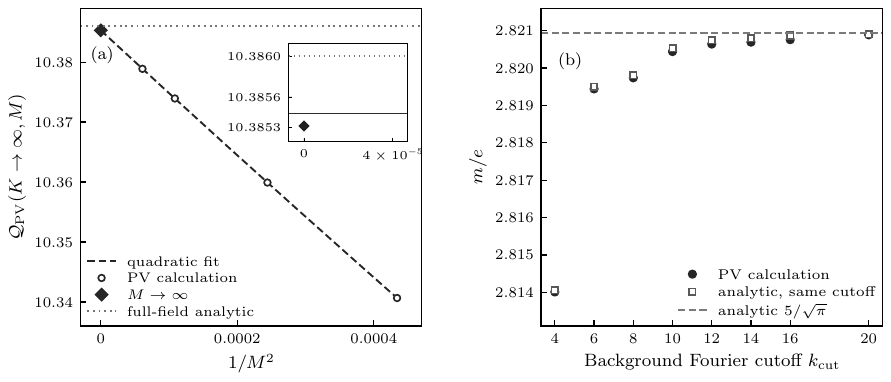}
\caption{Induced mass from the PV determinant. (a) Response versus $1/M^2$ at $k_{\rm cut}=16$: numerical points, quadratic fit (dashed), and extrapolation (diamond). The analytic full-field result (dotted) and truncated-field result (solid) are resolved in the inset. (b) Mass ratio versus background Fourier cutoff: PV results (circles), analytic values for the same truncated background (squares), and the full-field value $5/\sqrt\pi$ (dashed). The $k_{\rm cut}=20$ point uses the extended spectral ranges $K\le500$, $M\le128$; all other points use $K\le220$, $M\le64$.}
\label{fig:mass}
\end{figure*}

Figure~\ref{fig:mass}(a) shows the PV-mass extrapolation at $k_{\rm cut}=16$. The inset resolves the full analytic value from the value for the truncated background. Figure~\ref{fig:mass}(b) compares the numerical masses with both analytic references as $k_{\rm cut}$ increases. With the largest spectral range, $K\le500$ and $M\le128$, we find $m/e=2.82088882$ at $k_{\rm cut}=20$. The analytic value for the same truncated background is $2.82090677$, while the full continuum value is $2.82094792$. The difference between the first two numbers is approximately $1.8\times10^{-5}$. Changing the extrapolation fits gives $2.82085910\le m/e\le2.82089533$ at this background cutoff. The fit values all lie below the same-background analytic result, indicating a small one-sided bias. The spread of the fits is therefore not a rigorous error bound.

\section{Twelve-point function at $Q=1$}
\label{sec:qone}

\subsection{Exact continuum result}
\label{subsec:qone_exact}

The zero-flux determinant tests do not involve the generalized APS condition for boundary fermion zero modes~\cite{Witten:2026ezr}. We now test that condition in the $Q=1$ sector, using the twelve-point function defined in Eqs.~(\ref{eq:qone_operator}) and (\ref{eq:qone_ratio_definition}). Unlike a partition function without insertions, this correlator is nonzero and has a calculable dependence on the gauge field. Fermion-number-violating correlators in related two-dimensional chiral theories have previously been studied using overlap fermions~\cite{Kikukawa:1997dv}.

We choose periodic fluctuations depending only on $x$, which give a nonuniform field strength while keeping the Landau matrix block diagonal, as noted in Sec.~\ref{subsec:qone_setup}. The transverse fluctuation $\varphi$ and the gauge function $\chi$ are
\begin{align}
\varphi(x)&=a\cos(2\pi x)+b\cos(4\pi x),\nonumber\\
\chi(x)&=c\cos(2\pi x).
\label{eq:qone_profiles}
\end{align}
The reference field corresponds to $a=b=c=0$. We compare three backgrounds,
\begin{align}
\mathrm O:&\quad (a,b,c)=(0.025,0,0.025),\nonumber\\
\mathrm A:&\quad (a,b,c)=(0.040,0,0.025),\nonumber\\
\mathrm B:&\quad (a,b,c)=(0.025,0.004,0.025).
\label{eq:qone_cases}
\end{align}
The three choices have different field strengths and are therefore not related by gauge transformations. The gauge function $\chi$ allows us to test a nonzero local phase and check the gauge dependence of the regulated magnitude.

To calculate the exact reference-normalized correlator, we combine the contribution of the zero-mode wave functions with the determinant over nonzero modes~\cite{Sachs:1991en}. For a positive charge $q$, let $f_i^{(q)}$ ($i=1,\ldots,q$) be an orthonormal basis of zero modes in the uniform reference field. The unnormalized zero modes in the target field are
\begin{equation}
e^{q\varphi-iq\chi}f_i^{(q)}(x,y).
\label{eq:qone_dressing}
\end{equation}
Their inner products form the $q\times q$ Gram matrix
\begin{equation}
G^{(q)}_{ij}=\int_{T^2}d^2x\,
 e^{2q\varphi}\overline{f_i^{(q)}}f_j^{(q)}.
\label{eq:qone_gram}
\end{equation}
The factor $(\det G^{(q)})^{-1/2}$ from orthonormalizing the zero-mode Slater determinant cancels the corresponding $(\det G^{(q)})^{1/2}$ in the primed determinant magnitude. The orthonormal zero-mode frame also carries a phase. It is determined by the reference-to-target overlap matrix
\begin{equation}
\Omega^{(q)}_{ij}=\int_{T^2}d^2x\,
 \overline{f_i^{(q)}}e^{q\varphi-iq\chi}f_j^{(q)}.
\label{eq:qone_overlap}
\end{equation}
After orthonormalization of the target zero modes, the overlap matrix is $\Omega^{(q)}(G^{(q)})^{-1/2}$. The target frame is aligned with the reference frame by right multiplication with the adjoint of its unitary polar factor. Since $G^{(q)}$ is positive definite, the determinant phase of the overlap is $\arg\det\Omega^{(q)}$. Let $r_j^{(3)}=x_j$, $r_j^{(4)}=y_j$, and $r_j^{(5)}=z_j$. With this alignment, the phase of the target-to-reference Slater determinant ratio $\mathcal S_q$ for a left-handed charge $q$ is
\begin{equation}
\arg\mathcal S_q=-q\sum_{j=1}^q\chi(r_j^{(q)})
                      -\arg\det\Omega^{(q)}.
\label{eq:qone_slater_frame}
\end{equation}
The right-handed contribution is complex conjugated. After combining the charges, the phase that the eta factor must cancel is therefore fixed by the two-dimensional zero modes,
\begin{equation}
\Phi_{\rm frame}=-\arg\det\Omega^{(3)}
 -\arg\det\Omega^{(4)}+\arg\det\Omega^{(5)}.
\label{eq:qone_frame_phase}
\end{equation}
We denote the required APS phase by $\Theta_{\rm req}=-\Phi_{\rm frame}$.
This frame phase arises from the alignment of the two zero-mode frames and is independent of the twelve insertion points. The primed determinant magnitude obeys
\begin{align}
\log\frac{|\det'\mathcal D_{L,q}[A]|}
              {|\det'\mathcal D_{L,q}[A^{\rm ref}]|}
={}&\frac12\log\det G^{(q)}\nonumber\\
&-\frac{q^2}{4\pi}\int_{T^2}d^2x\,|\nabla\varphi|^2.
\label{eq:qone_primed_exact}
\end{align}
We use this formula to check the numerical calculation of the nonzero-mode PV magnitude.

Combining the three charged fermions gives the exact complex ratio
\begin{align}
\log|R_1^{\rm exact}|={}&
 \sum_{q=3,4,5}q\sum_{j=1}^{q}
     \varphi(r_j^{(q)})\nonumber\\
 &-\frac{50}{4\pi}\int_{T^2}d^2x\,|\nabla\varphi|^2,
\label{eq:qone_exact_magnitude}\\
\arg R_1^{\rm exact}={}&
 -3\sum_{j=1}^3\chi(x_j)
 -4\sum_{j=1}^4\chi(y_j)\nonumber\\
 &+5\sum_{j=1}^5\chi(z_j).
\label{eq:qone_exact_phase}
\end{align}
The coefficient $50=3^2+4^2+5^2$ is the sum of squared charges of the three gauge-coupled fermions. For the profiles in Eq.~(\ref{eq:qone_profiles}), the integral is $\int|\nabla\varphi|^2=2\pi^2(a^2+4b^2)$. The second term in Eq.~(\ref{eq:qone_exact_magnitude}) has the same coefficient as the nonflat determinant at $Q=0$, while the first comes from the zero-mode insertions. The exact phase in Eq.~(\ref{eq:qone_exact_phase}) is entirely the local gauge phase at the insertion points, typically $0.1$--$0.3$ rad. It has no contribution from $\varphi$. The eta invariant does not generate this phase. Instead, we compare the computed APS phase $\Theta_{\rm APS}$ [defined below Eq.~(\ref{eq:qone_numerical_phase})] with $\Theta_{\rm req}$ and vary the zero-mode identification and bulk interpolation.

\begin{figure*}[!t]
\includegraphics[width=0.90\textwidth]{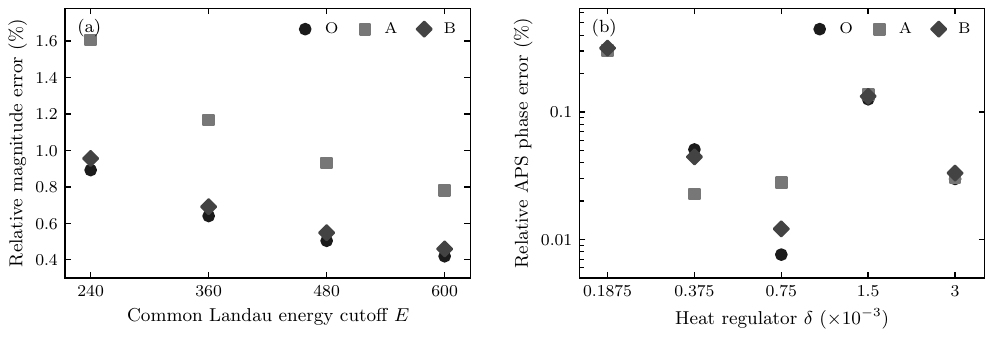}
\caption{$Q=1$ twelve-point correlator compared with the exact continuum result. (a) Relative magnitude errors at $M=24$, for the three nonflat backgrounds O, A, and B, as the common Landau energy cutoff $E$ increases. (b) Relative APS phase errors $100|\Theta_{\rm APS}/\Theta_{\rm req}-1|$ (percent) as a function of $\delta$ at $E_\eta=120$ and $p=24$.}
\label{fig:qone_checks}
\end{figure*}

\subsection{Magnitude and insertion points}

We compute the nonzero singular values and the zero-mode Slater determinants in the Landau basis of Sec.~\ref{subsec:qone_setup}. With the common Landau cutoff $E$, the number of retained levels depends on the charge according to Eq.~(\ref{eq:qone_energy_cutoff}). The PV mass is $M=24$. The numerical magnitude of $R_1$ is the product of the primed PV ratio and the three Slater-determinant ratios.

\begin{table}[t]
\begin{ruledtabular}
\begin{tabular}{ccccc}
$E$ & Landau levels & O & A & B\\
\hline
240 & $(80,60,48)$    & $0.892$ & $1.606$ & $0.956$\\
360 & $(120,90,72)$   & $0.640$ & $1.167$ & $0.691$\\
480 & $(160,120,96)$  & $0.504$ & $0.930$ & $0.548$\\
600 & $(200,150,120)$ & $0.419$ & $0.782$ & $0.459$\\
600, $\chi=0$ & $(200,150,120)$ & $0.231$ & $0.593$ & $0.272$
\end{tabular}
\end{ruledtabular}
\caption{Relative errors (percent) of $|R_1|$ for the three physically distinct $Q=1$ backgrounds. The Landau cutoff $E$ is common to all charges, and $M=24$. The second column lists the level bounds $n_{\rm max}=\lfloor E/q\rfloor$ for $q=3,4,5$, in that order. All results use insertion-point set 1 of Table~\ref{tab:qone_positions}. The last row repeats the $E=600$ calculation with $c=0$, i.e., in the gauge $\chi=0$.}
\label{tab:qone_cutoffs}
\end{table}

Table~\ref{tab:qone_cutoffs} and Fig.~\ref{fig:qone_checks}(a) show that the magnitude error decreases steadily in all three backgrounds as $E$ is raised. At the largest tested cutoff all three errors are below one percent. The residual differences include finite-$E$ and finite-$M$ effects.

The exact magnitude does not depend on $\chi$. In the gauge $\chi=0$, the errors at $E=600$, $M=24$ fall to $0.23$--$0.59\%$ (last row of Table~\ref{tab:qone_cutoffs}). The pure-gauge part accounts for about $0.19$ percentage points of the error in each background. The zero-mode Slater magnitudes agree between the two gauges within $5\times10^{-14}$ in their logarithms, so the difference comes from the truncated nonzero-mode PV spectrum. Indeed, a pure-gauge deformation alone, $\varphi=0$ and $\chi=0.025\cos(2\pi x)$, changes the PV magnitude by $0.187\%$, although the exact change is zero. This artifact falls from $0.437\%$ to $0.187\%$ as $E$ increases from $240$ to $600$. Halving the amplitude of $\chi$ at $E=600$ reduces it to $0.047\%$ and doubling it raises it to $0.751\%$, consistent with a quadratic dependence on the amplitude. Being even in $\chi$ at leading order, the artifact does not show up in a comparison of $\chi$ with $-\chi$.

The insertion points can be varied without changing the gauge background. Table~\ref{tab:qone_positions} shows four choices for background B. The point sets are specified in Appendix~\ref{app:qone}. The analytic and numerical correlator ratios change appreciably with position, both in magnitude and phase. The ratio of results at two different point sets agrees with the analytic local dressing of Eq.~(\ref{eq:qone_dressing}) to better than $10^{-12}$ at the largest cutoff. In this positional ratio the common PV and eta factors cancel. This comparison therefore tests the zero-mode wave functions and their position dependence, not the eta invariant.

\begin{table}[t]
\begin{ruledtabular}
\begin{tabular}{ccccc}
Set & $|R_1^{\rm exact}|$ & $|R_1^{\rm num}|$
& $\arg R_1^{\rm exact}$ & $\arg R_1^{\rm num}$\\
\hline
1 & $1.04315$ & $1.04794$ & $+0.263915477$ & $+0.263915698$\\
2 & $0.82685$ & $0.83065$ & $+0.299613560$ & $+0.299613781$\\
3 & $1.42992$ & $1.43648$ & $+0.161623347$ & $+0.161623568$\\
4 & $0.55148$ & $0.55401$ & $-0.127320361$ & $-0.127320140$
\end{tabular}
\end{ruledtabular}
\caption{Complex twelve-point ratio for background B with four insertion-point sets. The two-dimensional Landau cutoff is $E=600$ and $M=24$. At fixed background the PV and eta factors are the same for all four sets; the insertion factors give the variation. The eta phase uses the three-dimensional data at $E_{\eta}=120$, $p=24$, $\delta=0.00075$. Angles are principal phases in radians.}
\label{tab:qone_positions}
\end{table}

\subsection{APS phase and consistency}
\label{subsec:qone_aps}

We next test the phase, for which the three-dimensional calculation is essential. The numerical correlator is assembled from the Slater phases and the cylinder eta invariant,
\begin{align}
\arg R_1^{\rm num}
={}&\arg{\cal S}_3+\arg{\cal S}_4-\arg{\cal S}_5\nonumber\\
&-\pi(\Delta\eta_3+\Delta\eta_4-\Delta\eta_5),
\label{eq:qone_numerical_phase}
\end{align}
where $\mathcal{S}_q$ is the numerical target-to-reference Slater ratio and $\Delta\eta_q$ is the difference between the three-dimensional eta invariants for charge $q$ of the target and uniform-reference cylinders, evaluated with the same gluing map and spectral cutoffs. The negative signs for $q=5$ implement the right-handed conjugation. We denote the combined APS phase by $\Theta_{\rm APS}=-\pi(\Delta\eta_3+\Delta\eta_4-\Delta\eta_5)$. We use a common three-dimensional Landau energy bound $E_{\eta}=120$, polynomial basis size $p=24$ along the cylinder, and heat-kernel regulator $\delta=0.00075$ unless stated otherwise. These cutoffs are separate from those of the two-dimensional magnitude calculation.

\begin{table}[t]
\begin{ruledtabular}
\begin{tabular}{lrrr}
 & O & A & B\\
\hline
$\Theta_{\rm req}$ (mrad) & $1.887195$ & $3.048992$ & $1.822135$\\
\hline
\multicolumn{4}{l}{Signed residuals ($\mu$rad)}\\
$\delta=0.0001875$ & $-5.818$ & $-9.208$ & $-5.769$\\
$\delta=0.000375$ & $-0.959$ & $-0.693$ & $-0.809$\\
$\delta=0.00075$ & $+0.144$ & $+0.849$ & $+0.221$\\
$\delta=0.0015$ & $+2.373$ & $+4.173$ & $+2.417$\\
$\delta=0.003$ & $+0.563$ & $+0.930$ & $+0.605$
\end{tabular}
\end{ruledtabular}
\caption{Required APS phases $\Theta_{\rm req}$ (mrad) and signed residuals $\Theta_{\rm APS}-\Theta_{\rm req}$ ($\mu$rad). The required phases follow from the two-dimensional overlap matrices in Eq.~(\ref{eq:qone_overlap}) via Eq.~(\ref{eq:qone_frame_phase}). The cylinder calculation uses $E_\eta=120$ and $p=24$.}
\label{tab:qone_eta}
\end{table}

The overlap matrices in Eq.~(\ref{eq:qone_overlap}) determine the required APS phase. In background O, for example, the phases of $\det\Omega^{(q)}$ are $-5.455$, $-10.881$, and $-18.223$ mrad for $q=3,4,5$. The required APS phase is $+1.887$ mrad. The single-charge eta phases can in addition contain a term proportional to $q^2$, which cancels in the anomaly-free combination. The bulk-path comparison below tests this term.

Table~\ref{tab:qone_eta} compares the phase obtained from the cylinder spectrum with this target. At $\delta=0.00075$, the APS phase agrees with the required value within $0.03\%$ for all three backgrounds, with absolute residuals below $10^{-6}$ rad. Increasing $p$ from 18 to 24 at fixed $E_\eta=120$ and $\delta=0.00075$ reduces the largest relative error from $0.058\%$ to $0.028\%$. Figure~\ref{fig:qone_checks}(b) shows that the dependence on the heat regulator remains nonmonotonic. This comparison isolates the contribution of the three-dimensional eta invariant from the larger local phase of the correlator. The resulting twelve-point ratios at the four insertion-point sets in Table~\ref{tab:qone_positions} have phase errors of about $2.2\times10^{-7}$ rad.

The generalized APS condition identifies the zero-mode spaces at the two ends of the cylinder using a unitary map $U$. The final correlator must be independent of this choice. To test this compensation directly, we keep the endpoint frames and the reference normalization fixed and vary the gluing map. When all eigenphases of $V$ are shifted by $0.23$ rad for each charge, the combined zero-mode phase changes by $(3+4-5)\times0.23=0.46$ rad, while the directly computed eta phase changes by about $-0.460000049$ rad. The residual is $-4.89\times10^{-8}$ rad at $E_\eta=120$, $p=24$, and $\delta=0.003$. For a nonuniform determinant-preserving diagonal change of $V$, it is $1.89\times10^{-7}$ rad, as shown in Appendix~\ref{app:qone}.

We also compare two ordered interpolations in the cylinder with identical boundary fields. We turn on the first harmonic $a\cos(2\pi x)$ of $\varphi$ before the remaining deformation $b\cos(4\pi x)$ and $\chi$, or in the reverse order. The APS index theorem~\cite{Atiyah:1975jf,Alvarez-Gaume:1984zst} gives a quantitative benchmark for the difference between the two ordered paths. In our orientation, the relative Abelian Chern--Simons integral is $-4\pi^2ac$ (Appendix~\ref{app:qone}), so the phase difference between the two paths for a single charge $q$ in the left-handed convention is
\begin{equation}
\Delta\Theta_q^{\rm exact}=q^2\pi ac.
\label{eq:qone_bulk_exact}
\end{equation}
For background B this is $q^2\times1.963495$ mrad, independent of $b$. Table~\ref{tab:qone_bulk_charge} compares the numerical values directly with this anomaly coefficient. At fixed $p=24$ and $\delta=0.003$, the charge-resolved coefficients reproduce about $96\%$ of the exact value at $E_\eta=120$ and $97.6$--$97.7\%$ at $E_\eta=180$. The reproduced fraction is nearly the same for the three charges. In each case the combination $3^2+4^2-5^2=0$ strongly suppresses the bulk-path dependence. At $E_\eta=180$ the combined phase difference is only $0.01665$ mrad in absolute value. This calculation tests both the expected $q^2$ dependence and its absolute coefficient.

\begin{table}[t]
\begin{ruledtabular}
\begin{tabular}{crrr}
Charge & Exact & $E_\eta=120$ & $E_\eta=180$\\
\hline
$3$ & $17.671$ & $16.991$ & $17.251$\\
$4$ & $31.416$ & $30.221$ & $30.678$\\
$5$ & $49.087$ & $47.241$ & $47.946$\\
$3+4-5$ & $0$ & $-0.03029$ & $-0.01665$
\end{tabular}
\end{ruledtabular}
\caption{Charge-resolved phase difference between the two ordered bulk interpolations in background B. The exact target follows from Eq.~(\ref{eq:qone_bulk_exact}); APS results use $p=24$ and $\delta=0.003$. The individual phases are given in the left-handed convention. The last row combines charges with signs $+,+,-$ before rounding the individual entries. All entries are in mrad.}
\label{tab:qone_bulk_charge}
\end{table}

The accuracy of the three-dimensional calculation depends on both the spectral cutoffs and the heat regulator. The regulator giving the best agreement need not be the same for different tests. We quote the cutoffs used in each test and do not claim a common spectral limit.

Finally, changing $\chi\to-\chi$ reverses the exact gauge phase. The numerically computed change agrees with it within $5.5\times10^{-7}$ rad. Details of these checks are given in Appendix~\ref{app:qone}.

\subsection{Link-interpolated gauge fields}
\label{subsec:qone_links}

We now apply the $Q=1$ spectral calculation to gauge fields reconstructed from lattice links. We construct the $6\times6$ link holonomies of backgrounds O, A, and B in Eq.~(\ref{eq:qone_cases}) and interpolate them using the procedure of Sec.~\ref{subsec:qone_setup}, with Eq.~(\ref{eq:interpolation}) and the smoothing width $d=0.05$. The reconstructed fields carry one unit of flux and reproduce the original link holonomies.

The transverse norm $\int d^2x\,|\nabla\varphi|^2$ of the reconstructed field is $86.0\%$, $86.0\%$, and $83.4\%$ of its original continuum value for O, A, and B, respectively. The reduction comes from sampling the field on six sites and from the cell profile $w$. Modes with $|k_x|\le2$ and $k_y=0$ carry $98.6$--$99.2\%$ of this norm. Higher $x$ harmonics contribute $0.6$--$1.1\%$ and $y$-dependent modes $0.2$--$0.3\%$.

For the spectral test, we consider the reconstructed background O, retaining the couplings between magnetic guiding-center sectors. The PV calculation and zero-mode contributions give a correlator magnitude that differs from the analytic result for the same Fourier-truncated field by $0.214\%$ at $E=600$, $M=24$, and $k_{\rm cut}=24$. Table~\ref{tab:qone_link_eta} compares the computed APS phase with the required phase from independent two-dimensional zero-mode overlaps. The latter is $1.583693$ mrad for the same Fourier-truncated background. We fix $p=12$ and $k_{\rm cut}=24$ and vary $E_\eta$ and $\delta$.

\begin{table}[t]
\begin{ruledtabular}
\begin{tabular}{cccc}
$E_\eta$ & $\delta=0.00075$ & $\delta=0.0015$ & $\delta=0.003$\\
\hline
200 & 1.655501 & 1.599186 & 1.582130\\
240 & 1.260645 & 1.556173 & 1.581802\\
300 & 1.441424 & 1.585391 & 1.582289\\
Required & 1.583693 & 1.583693 & 1.583693
\end{tabular}
\end{ruledtabular}
\caption{Computed ($\Theta_{\rm APS}$) and required ($\Theta_{\rm req}$) APS phases for the link-interpolated background O, in mrad. The required phase is obtained from the two-dimensional zero-mode overlaps of the same Fourier-truncated gauge field. The APS calculation uses $p=12$ and $k_{\rm cut}=24$.}
\label{tab:qone_link_eta}
\end{table}

At $\delta=0.003$, the computed APS phase agrees with the required value within $0.12\%$ at all three Landau cutoffs. The variation among these results is below $5\times10^{-7}$ rad. This shows that the generalized APS prescription can also be evaluated accurately for a gauge field reconstructed from lattice links, including couplings between different magnetic guiding-center sectors. At smaller values of $\delta$, the cutoff dependence remains nonmonotonic. A systematic study of the joint spectral and heat-kernel limits remains necessary.

\section{Summary and discussion}
\label{sec:discussion}

We have tested a prescription for continuum chiral fermions coupled to interpolated lattice gauge fields in the two-dimensional $3450$ theory. We calculated the determinant magnitude using PV regularization and its phase from a three-dimensional APS spectrum, then compared their product with exact continuum results.

The first main result is that, in the $Q=0$ sector, the complex weight reproduces a phase-sensitive Polyakov loop. Its value on the $16\times16$ toron grid differs from the analytic result by only $0.034\%$, with common spectral cutoffs over the grid. Dropping the APS phase instead changes the Polyakov loop from about $0.804$ to $0.025$. Cutoff scans on fixed toron sets show that the phases improve with the radial and angular cutoffs, although their dependence on the heat regulator is not monotonic.

The nonflat determinant ratios agree with the exact results within $0.4\%$ for the two weak-curvature $3\times3$ backgrounds and within $0.5\%$ for the $4\times4$ and $6\times6$ examples. The errors grow to $1$--$2\%$ for the two stronger-curvature backgrounds. The relative APS phase of configuration II approaches the analytic zero as the angular cutoff increases. That of configuration I behaves less regularly. The vacuum-polarization calculation gives the induced mass $m/e=2.82089$, close to $5/\sqrt\pi$. This result tests the quadratic effective action, complementing the tests of the complex determinant.

The second main result concerns $Q=1$, where the generalized APS boundary condition links zero-mode spaces at the ends of a cylinder. For three smooth nonflat backgrounds, the twelve-point correlator magnitude agrees with the exact result within $0.8\%$. In the gauge $\chi=0$ the errors are $0.23$--$0.59\%$. The difference quantifies the gauge dependence introduced by the truncated Landau basis.

While the magnitude is determined entirely by the two-dimensional spectral calculation, the phase provides a direct test of the three-dimensional eta invariant. Two-dimensional zero-mode overlaps fix the required phase, and the cylinder spectrum reproduces it within $0.03\%$ for all three smooth backgrounds at common cutoffs. The resulting twelve-point correlator agrees with the exact phase within $10^{-6}$ rad. The single-charge phase differences reproduce about $98\%$ of the Chern--Simons prediction $q^2\pi ac$, while the anomaly-free combination suppresses the phase difference to below $0.02$ mrad. The changes of the eta phase and the zero-mode phase induced by varying the gluing map $V$ also cancel to high numerical accuracy. Together these checks support the numerical implementation of the generalized APS prescription~\cite{Witten:2026ezr}.

Some of these observations are not specific to the backgrounds studied here. Combining the positive quadratic form with first-order Dirac matrix elements retains the sign information required by the APS trace. The heat regulator must be balanced against the spectral cutoffs. Truncating the PV basis produces a measurable gauge artifact. Finally, the average phase factor $0.285$ in the flat-toron integral indicates sizable phase cancellations in direct complex-weight sampling~\cite{Witten:2026ezr}, unlike dual formulations of this model without a sign problem \cite{Berkowitz:2023pnz}.

At zero flux, our tests include the flat-toron integral and selected nonflat lattice backgrounds. At unit flux, the main twelve-point correlator tests use smooth continuum backgrounds. We have also constructed continuum fields from lattice links, examined their Fourier content, and applied the PV and APS spectral calculations to them. The APS phase for the link-interpolated background O agrees with the required value within $0.12\%$ at three Landau cutoffs for a common heat regulator. A systematic study of the joint spectral and heat-kernel limits remains necessary.

For the lattice backgrounds, the continuum benchmarks and numerical spectra use the same gauge-field interpolation, with the effects of the background Fourier truncation examined separately. These comparisons test the PV and APS prescriptions, but do not establish the continuum limit of the lattice theory or independence from the interpolation profile. A related four-dimensional analysis found profile dependence in a local counterterm~\cite{Liu:2026matching}; the corresponding dependence in this two-dimensional model remains to be studied.

\section*{Data Availability}
The numerical codes and data supporting this work are available as ancillary files accompanying the arXiv submission.

\begin{acknowledgments}
The author thanks Yudai Hamada for fruitful discussions about the present work and its numerical implementation. The author also thanks Henry Lamm for helpful discussions and comments on the manuscript.
This work was partially supported by the Japan Society for the Promotion of Science (JSPS) KAKENHI Grant Nos.~23K03425 and 22H05118. OpenAI ChatGPT (GPT-5 and GPT-6 model families) assisted with numerical code development, figure preparation and manuscript polishing under the author's direction. The author is responsible for independently validating the numerical results and their interpretation.
\end{acknowledgments}

\appendix

\section{Local interpolation}
\label{app:interp}

In the zero-flux sector the plaquette angles satisfy $\sum_{ij}\phi_{ij}=0$.  Real lifts of the link angles may be chosen so that
\begin{equation}
\theta^x_{ij}+\theta^y_{i+1,j}-\theta^x_{i,j+1}-\theta^y_{ij}
=\phi_{ij}.
\end{equation}
The functions $S$ and $w=S'$ in Sec.~\ref{sec:setup} are defined as follows. Let
\begin{equation}
f(u)=
\begin{cases}
0,& u\le0,\\
\displaystyle\frac{e^{-1/u}}{e^{-1/u}+e^{-1/(1-u)}},&
0<u<1,\\
1,&u\ge1.
\end{cases}
\end{equation}
With $d=0.05$, we take
\begin{align}
w(s)&=\frac{f((s-d)/d)\,f((1-d-s)/d)}{1-3d},
\nonumber\\
S(s)&=\int_0^s du\,w(u).
\label{eq:bump_details}
\end{align}
Thus $w$ vanishes smoothly near each cell boundary and is normalized to unity. We evaluate $S$ on a fine integration grid and the Fourier coefficients of $w$ by Gaussian quadrature. In each cell, Eq.~(\ref{eq:interpolation}) gives the link holonomies on its four sides, because $\int_0^1w=1$.  The values of the tangential connection agree on neighboring cell boundaries.  The connection can therefore be joined smoothly across the cell edges. This interpolation differs from smoothing a piecewise-defined connection by gradient flow, another possibility discussed by Witten~\cite{Witten:2026ezr}.

Differentiating Eq.~(\ref{eq:interpolation}) gives the curvature in Eq.~(\ref{eq:cell_curvature}).  For a lattice gauge transformation with compatible real lifts, the four vertex gauge parameters admit a smooth interpolation with factors $S(s)$ and $S(t)$.  Its derivative changes the connection by $A\mapsto A+d\Lambda$ and leaves $F$ unchanged.  This verifies gauge covariance for the interpolation in the Abelian case.  The construction also makes the dependence on the chosen smoothing function explicit.

The Fourier transform of the curvature, given explicitly in Eq.~(\ref{eq:Fk}), follows by integrating Eq.~(\ref{eq:cell_curvature}) over each cell. It determines the geometric normalization used in the mass calculation.

The two $3\times3$ nonflat backgrounds used in Table~\ref{tab:nonflat} are specified as follows. With rows labeled by $i$ and columns by $j$, the unscaled plaquette-angle arrays are
\begin{align}
\phi^{(\mathrm{I})}_{ij}&=\begin{pmatrix}
-0.480000&-0.049218& 0.225875\\
 0.346761& 0.346761& 0.105685\\
-0.155888&-0.049218&-0.290758
\end{pmatrix},\nonumber\\
\phi^{(\mathrm{II})}_{ij}&=\begin{pmatrix}
-0.044516& 0.540294& 0.540294\\
-0.493213& 0.149477&-0.158063\\
-0.800000&-0.274569& 0.540294
\end{pmatrix}.
\label{eq:shape_arrays}
\end{align}
The small deviations of the sums from zero arise from the displayed decimal rounding. In the calculation, the arrays have exactly zero sum. The shapes were constructed from points 31 and 17 (indexed from zero), respectively, of an unscrambled nine-dimensional Sobol sequence. We apply the inverse standard-normal distribution to the nine coordinates, subtract their mean, and normalize the largest absolute entry to $0.48$ for I and $0.80$ for II. This deterministic prescription specifies the unrounded arrays used in the calculations. For the four entries of Table~\ref{tab:nonflat}, configuration I is multiplied by $0.5$ or $0.9602898565$ and configuration II by $0.25$ or $0.5$. The factor $0.9602898565$ fixes the larger value of $\max_{i,j}|\phi_{ij}|$ for configuration I and is not a fitted parameter.

For the larger-lattice checks, let $N=4$ or $6$, with indices $i,j=0,\ldots,N-1$ and the same smoothing function $w$. We set $h_x=0.21875$, $h_y=0.28125$, and $F_{\mathrm{amp}}=1.8$. The link angles are
\begin{align}
\theta^x_{ij}&=\frac{2\pi h_x}{N},\nonumber\\
\theta^y_{ij}&=\frac{2\pi h_y}{N}
 +\frac{F_{\mathrm{amp}}}{2N^2\sin(\pi/N)}
  \sin\frac{2\pi(i-1/2)}{N}.
\label{eq:larger_lattice_links}
\end{align}
Their oriented plaquette angles are
\begin{equation}
\phi_{ij}=\frac{F_{\mathrm{amp}}}{N^2}\cos\frac{2\pi i}{N},
\qquad \sum_{i,j}\phi_{ij}=0.
\label{eq:larger_lattice_flux}
\end{equation}
Thus $\max_{i,j}|\phi_{ij}|=0.1125$ for $N=4$ and $0.0500$ for $N=6$.

\section{PV spectra at $Q=0$}
\label{app:pv}

\subsection{Flat torons}

For a flat connection, the positive eigenvalues entering the PV determinant of charge $q$ are
\begin{equation}
\lambda_n(q,h)=(2\pi)^2\sum_{\mu=x,y}
    (n_\mu+\tfrac12+qh_\mu)^2.
\label{eq:flat_spectrum}
\end{equation}
The normalized finite-cutoff determinant is
\begin{align}
\log\frac{W_{\rm PV}(h)}{W_{\rm PV}(0)}
={}&\frac12\sum_{q=3,4,5}\sum_{|n_\mu|\le K}\biggl[
\log\frac{\lambda_n(q,h)}{\lambda_n(0,0)}\nonumber\\
&-\log\frac{\lambda_n(q,h)+M^2}
               {\lambda_n(0,0)+M^2}\biggr].
\label{eq:flat_pv_product}
\end{align}
We use $K=256,384,512,768,1024,1536$ and $M=16,24,32,48,64,96$ for the 256 toron points. At each fixed $M$ we fit the four largest $K$ values quadratically in $1/K^2$. The four largest PV masses are then fitted quadratically in $1/M^2$. Changing the fit windows tests the extrapolation. As a further check, we integrate the proper-time heat trace of the same free spectrum. The two extrapolated logarithmic magnitudes differ by no more than $2.4\times10^{-6}$ over the toron grid.

\subsection{Quadratic response}

The PV calculation uses Fourier modes with $-K\le n_x,n_y\le K$ and antiperiodic fermion boundary conditions. For charge $q$, the free chiral operator is diagonal and is given by Eq.~(\ref{eq:D0}). For a small gauge background $tA$, we denote the Fourier matrix $\widetilde D_q$ by $D(t)=D_0+t\mathcal V$, where the matrix elements of $\mathcal V$ are obtained from the Fourier modes of $A_x,A_y$.  Each charged fermion contributes to the magnitude through
\begin{align}
\log\frac{W_{{\rm PV},q}(t)}{W_{{\rm PV},q}(0)}
={}&\frac12\operatorname{Tr}\log\frac{D(t)^\dagger D(t)}{D_0^\dagger D_0}
\nonumber\\[-1mm]
&-\frac12\operatorname{Tr}\log
\frac{D(t)^\dagger D(t)+M^2}{D_0^\dagger D_0+M^2}.
\label{eq:pv_trace}
\end{align}
The ratios in this equation denote differences of trace logarithms. They do not require the matrices in the numerator and denominator to commute.  The neutral fermion cancels in the normalized ratio.

The coefficient of $-t^2$ in Eq.~(\ref{eq:pv_trace}) is
\begin{align}
\mathcal Q_{K,M}^{(q)}={}&\frac12\operatorname{Re}\operatorname{Tr}
 \bigl[(D_0^{-1}\mathcal V)^2\bigr]
 +\frac12\operatorname{Tr}(\mathcal M_0^{-1}\mathcal V^\dagger\mathcal V)
\nonumber\\
&-\frac14\operatorname{Tr}\bigl[(\mathcal M_0^{-1}\mathcal M_1)^2\bigr],
\label{eq:pv_hessian}
\end{align}
where $\mathcal M_0=D_0^\dagger D_0+M^2$ and $\mathcal M_1=D_0^\dagger\mathcal V+\mathcal V^\dagger D_0$. We sum this expression over $q=3,4,5$ and evaluate the finite matrix trace in Eq.~(\ref{eq:pv_hessian}).

For the induced-mass calculation, the fermion Fourier cutoffs are $K=40,60,80,100,140,180,220$ and the PV masses are $M=8,12,16,24,32,48,64$. For the extended calculation, the fermion cutoffs are $K=100,140,180,220,260,320,400$, and in the largest data set $K=140,180,220,260,320,400,500$. We add the regulator masses $M=96,128$ in the largest data set. At each $M$ we fit the largest four available $K$ values linearly in $1/K^2$. The four largest PV masses then determine a quadratic fit in $1/M^2$. To assess the uncertainty of the extrapolation, we repeat the fits with three or five $K$ values, with five or six regulator masses, and with a cubic fit in $1/M^2$.

At fixed PV mass $M$, the infinite-volume continuum Fourier response provides an analytic benchmark, up to exponentially small finite-volume corrections on the torus. For a transverse background with Fourier momentum $P=2\pi\sqrt{k_x^2+k_y^2}$, it is
\begin{align}
\mathcal Q(M)&=\frac{25}{2\pi}\sum_{k\ne0}|\partial_t A_k^T|^2g_M(P),
\nonumber\\
g_M(P)&=\frac{4M^2}{P\sqrt{P^2+4M^2}}
\operatorname{artanh}\frac{P}{\sqrt{P^2+4M^2}}.
\label{eq:finite_M_response}
\end{align}
This expression approaches the unregularized polarization as $g_M(P)=1-P^2/(6M^2)+O(M^{-4})$ at fixed momentum. We evaluate the sum with the same truncated background Fourier modes as in the PV matrix calculation and compare it with the spectral $K\to\infty$ fits at $k_{\rm cut}=16$, where $\mathcal Q(M)$ rises from $10.2165$ at $M=24$ to $10.3790$ at $M=128$. The spectral values lie below the analytic ones by $1.0\times10^{-6}$, $4.5\times10^{-6}$, $9.2\times10^{-6}$, $2.8\times10^{-5}$, and $7.0\times10^{-5}$ at $M=24,48,64,96$, and $128$, respectively. The difference increases with $M$ because the finite-$K$ extrapolation becomes more demanding, as expected from the $M^2/K^2$ form of the leading cutoff correction. This comparison provides a check at finite regulator mass without relying on an extrapolation in $1/M^2$.

The same finite-$M$ response helps to interpret the nonflat determinant ratios in Table~\ref{tab:nonflat}. At quadratic order in the interpolated gauge field, the regulator action is
\begin{equation}
S_{\rm PV}^{(2)}(M)=\frac{25}{8\pi^3}
\sum_{\substack{k\ne0\\ |k_x|,|k_y|\le k_{\rm cut}}}
\frac{|F_k|^2}{k_x^2+k_y^2}
g_M(2\pi|k|).
\label{eq:nonflat_finite_M}
\end{equation}
Table~\ref{tab:nonflat_pv_budget} decomposes the logarithmic deviation of the PV determinant at $K=20$, $M=24$, and $k_{\rm cut}=6$ from the complete massless determinant into the background Fourier truncation, the quadratic finite-$M$ correction of Eq.~(\ref{eq:nonflat_finite_M}), and a remainder. The remainder contains both the finite fermion cutoff and terms beyond quadratic order in the massive regulator action, which this comparison does not separate. Most of it appears to be a finite-$K$ effect. At fixed $M=24$ it decreases with $K$ over the measured range $K=12,14,16,18,20$. For the first row of Table~\ref{tab:nonflat_pv_budget}, linear fits in $1/K^2$ using three to five of the largest values leave $(0.31$--$0.55)\times10^{-3}$ at $K\to\infty$, and including a $1/K^4$ term reduces this to about $0.073\times10^{-3}$. The five-point second-order polynomial fits shown in Fig.~\ref{fig:nonflat} give $K\to\infty$ intercepts of $(7.742,2.094,1.543,6.172)\times10^{-3}$ in the order of Table~\ref{tab:nonflat_pv_budget}, against $(7.669,2.080,1.518,6.074)\times10^{-3}$ for the sum of the background truncation and the quadratic finite-$M$ term. Increasing $K$ from $20$ to $24$ at fixed $M=24$ reduces the remainder from $1.367\times10^{-3}$ to $0.959\times10^{-3}$ for configuration I with maximum plaquette angle $0.240$, and from $1.102\times10^{-3}$ to $0.779\times10^{-3}$ for configuration II with maximum angle $0.200$.

\begin{table}[tbp]
\begin{ruledtabular}
\begin{tabular}{lrrrr}
Background & Total & Truncation & Finite $M$ & Remainder\\
\hline
I, $0.461$ & $12.657$ & $0.415$ & $7.254$ & $4.988$\\
I, $0.240$ & $3.446$ & $0.113$ & $1.967$ & $1.367$\\
II, $0.200$ & $2.621$ & $0.087$ & $1.431$ & $1.102$\\
II, $0.400$ & $10.484$ & $0.349$ & $5.725$ & $4.409$
\end{tabular}
\end{ruledtabular}
\caption{Contributions to the logarithmic PV magnitude discrepancy for the four $3\times3$ nonflat backgrounds. All entries are multiplied by $10^3$. The last column includes finite-$K$ and higher-order regulator effects.}
\label{tab:nonflat_pv_budget}
\end{table}

Table~\ref{tab:mass} lists the extrapolated mass for each background cutoff, and Table~\ref{tab:mass_extended} shows how it changes as the fermion and regulator ranges are extended at $k_{\rm cut}=12$ and $16$.

\begin{table}[tbp]
\begin{ruledtabular}
\begin{tabular}{cccc}
$k_{\rm cut}$ & $c_{\rm PV}$ & $(m/e)_{\rm PV}$ & $(m/e)_{\rm cut}$\\
\hline
$4$  & $7.91863355$ & $2.81400667$ & $2.81404742$\\
$6$  & $7.94924479$ & $2.81944051$ & $2.81949887$\\
$8$  & $7.95092296$ & $2.81973810$ & $2.81980135$\\
$10$ & $7.95488542$ & $2.82044064$ & $2.82052360$\\
$12$ & $7.95601542$ & $2.82064096$ & $2.82073834$\\
$14$ & $7.95630233$ & $2.82069182$ & $2.82079407$\\
$16$ & $7.95667621$ & $2.82075809$ & $2.82087083$\\
$20$ & $7.95741374$ & $2.82088882$ & $2.82090677$\\
Full & $25/\pi$ & $2.82094792$ & $2.82094792$
\end{tabular}
\end{ruledtabular}
\caption{Induced mass from the PV spectral calculation. All numerical rows except $k_{\rm cut}=20$ use $K\le220$, $M\le64$; the $k_{\rm cut}=20$ row uses the extended ranges $K\le500$, $M\le128$. The last column is the analytic value for the same Fourier-truncated background.}
\label{tab:mass}
\end{table}

\begin{table}[tbp]
\begin{ruledtabular}
\begin{tabular}{cccc}
$K_{\max}$ & $M_{\max}$ & $m/e$ ($k_{\rm cut}=12$) & $m/e$ ($k_{\rm cut}=16$)\\
\hline
$220$ & $64$ & $2.82064096$ & $2.82075809$\\
$260$ & $64$ & $2.82066328$ & $2.82078042$\\
$320$ & $64$ & $2.82067097$ & $2.82078811$\\
$400$ & $64$ & $2.82067423$ & $2.82079136$\\
$500$ & $64$ & $2.82067564$ & $2.82079278$\\
$500$ & $96$ & $2.82072120$ & $2.82084894$\\
$400$ & $128$ & $2.82071051$ & $2.82084155$\\
$500$ & $128$ & $2.82072291$ & $2.82085395$
\end{tabular}
\end{ruledtabular}
\caption{Induced mass at the extended fermion and regulator cutoffs. Each entry uses four-point fits in $1/K^2$ and $1/M^2$.}
\label{tab:mass_extended}
\end{table}

At $K_{\max}=400$ and $k_{\rm cut}=16$, extending the quadratic fit from the four largest PV masses ($M=24,32,48,64$) to six ($M=12,16,24,32,48,64$) lowers $m/e$ from $2.82079136$ to $2.82049706$, a shift of $2.94\times10^{-4}$. This larger shift comes from including the two lowest masses in the asymptotic $1/M^2$ fit. It is a sensitivity test, not an error estimate for the fits restricted to the larger masses. At $(K_{\max},M_{\max})=(500,128)$ and the same $k_{\rm cut}=16$, changing the fit windows and polynomial degree gives $2.82082706\le m/e\le2.82086016$. This spread measures the dependence on the extrapolation. Direct finite-amplitude singular-value calculations at smaller cutoffs also agree with the quadratic trace expansion at small $t$.

The transverse norm is obtained from the curvature alone,
\begin{equation}
I=\sum_{k\in\mathbb Z^2\setminus\{0\}}
\frac{|\partial_tF_k|^2}{4\pi^2(k_x^2+k_y^2)}.
\end{equation}
Here $F_k$ is the Fourier coefficient of the interpolated field strength. Restricting the sum to $|k_x|,|k_y|\le k_{\rm cut}$ defines $I_{k_{\rm cut}}$ in Eq.~(\ref{eq:mass_finite_background}). We find
\begin{align}
I_{12}&=2.60990008224,\nonumber\\
I_{16}&=2.61014526116,\nonumber\\
I_{20}&=2.61021176775,\nonumber\\
I&=2.61028791816.
\end{align}
Increasing the Fourier range used for the last number changes $I$ by less than $4\times10^{-9}$. The finite-$k_{\rm cut}$ difference is a property of the background representation, not an uncertainty in the numerical evaluation of the full norm.

\section{APS spectra at $Q=0$}
\label{app:aps}

We use $Y=D^2\times S^1_y$ with coordinates $(r,\theta,y)$, $0\le r\le1$, $0\le\theta<2\pi$, and $0\le y<1$. The $x$ cycle of the boundary torus is filled by the disk. The $y$ cycle remains noncontractible. The metric is
\begin{equation}
ds^2=dy^2+\frac{1}{(2\pi)^2}
\left[dr^2+\rho(r)^2d\theta^2\right],
\label{eq:aps_metric}
\end{equation}
where $\rho(r)=[1-s(r)]r+s(r)$. Here $s(r)=f((r-0.68)/0.24)$, with the smooth step $f$ defined in Appendix~\ref{app:interp}. Thus $s(r)=0$ for $r\le0.68$ and $s(r)=1$ for $r\ge0.92$, so $\rho=r$ near the disk center and $\rho=1$ near the boundary. The metric has a product form near the boundary. Both boundary cycles have unit physical length. The antiperiodic spin structure around the contractible $x$ cycle extends smoothly over the disk. In the spectral calculation we express eigenvalues of $H_Y$ in units of $2\pi$. Thus $\delta$ multiplies its squared eigenvalues, which are measured in Fourier-mode units. The eta invariant itself is unchanged by this positive rescaling.

For a flat boundary connection, we use
\begin{align}
A&=2\pi h_y\,dy+h_x f_{\rm flat}(r)\,d\theta,\nonumber\\
f_{\rm flat}(r)&=[1-s(r)]r^2+s(r).
\label{eq:flat_bulk_profile}
\end{align}
This connection is regular at the disk center and is constant in the radial direction near the boundary. All the flat-toron results use this same filling. The boundary operator is diagonalized separately in each Fourier sector labeled by $\ell\in\mathbb Z+\tfrac12$ along $\theta$ and $m+\tfrac12$ along $y$. After rescaling the radial spinor by the square root of the metric density, we use Jacobi-polynomial basis functions proportional to
\begin{equation}
r^{|\ell\mp1/2|+1/2}
P_j^{(0,|\ell\mp1/2|)}(2r^2-1),
\quad j=0,\ldots,p-1,
\label{eq:aps_radial_basis}
\end{equation}
for the two spin components. The negative-eigenvalue subspace of the boundary Dirac operator is removed by the APS projector. Before this boundary constraint, a flat Fourier sector has $2p$ radial degrees of freedom. Imposing it leaves $2p-1$. We retain modes with $-K_{\rm APS}\le\ell<K_{\rm APS}$ and $-K_{\rm APS}\le m+1/2+qh_y<K_{\rm APS}$, where $m\in\mathbb Z$. The boundary Fourier window contains $(2K_{\rm APS})^2$ sectors per charge. For nonflat fields the angular sectors mix, and the corresponding boundary projector is applied to the coupled matrix.

We evaluate the spectral asymmetry with the estimator of Eq.~(\ref{eq:eta_estimator_main}). The eigenvalues $\mu_j$ of the positive quadratic form of $H_Y^2$ are combined with the diagonal matrix elements of $H_Y$ in that eigenbasis. This follows the spectral strategy described by Witten~\cite{Witten:2026ezr}. The sign information of the first-order Dirac operator is needed in addition to the spectrum of its square.

In the finite basis, one must distinguish the projection of the first-order Dirac operator from the quadratic form of its square. If $P$ is the finite-basis projector, in general
\begin{equation}
(PH_Y P)^2
\ne PH_Y^2P.
\label{eq:projection_diff}
\end{equation}
We use the quadratic-form representation, in which $PH_Y^2P$ denotes the matrix with elements $\langle H_Yu,H_Yv\rangle$ for orthonormal basis vectors $u,v$. The heat trace takes the form
\begin{align}
\operatorname{Tr}\bigl[&
(PH_Y P)(PH_Y^2P)^{-1/2}
\nonumber\\
&\times\operatorname{erfc}\!\left(\sqrt{\delta PH_Y^2P}\right)
\bigr],
\label{eq:form_eta}
\end{align}
where the inverse square root and trace are evaluated on the finite spectral space, with the appropriate treatment of numerical zero modes. The operator used numerically has the opposite orientation to the convention of Eq.~(\ref{eq:weight}). Its eta invariant is converted accordingly.

Using the full matrix trace, Eq.~(\ref{eq:form_eta}), avoids assigning an independent sign to each eigenvector of a nearly degenerate $H_Y^2$ eigenspace. In the continuum definition the zero-eigenvalue contribution is included according to Witten's APS convention. The backgrounds reported here have no detected exact bulk zero modes, and zero-mode crossings in general backgrounds are outside the present tests.

\subsection{Flat torons}

For flat torons the boundary Fourier cutoff $K_{\rm APS}$ applies in both torus directions. The Fourier windows give two reflection identities when they preserve the following transformations. Under $h_x\to-h_x$, relabeling $\ell\to-\ell$ and a spin rotation with $\sigma_2$ sends the bulk Dirac operator to its negative. Under $h_y\to-h_y$, the corresponding operations use $m+1/2\to-(m+1/2)$ and $\sigma_3$. Both preserve the APS boundary projector. We denote the APS phase in the convention of Eq.~(\ref{eq:weight}) by $\theta_{\rm APS}$. Thus, for the toron values used in our reflection completion, at the same finite cutoffs,
\begin{align}
\theta_{\rm APS}(-h_x,h_y)
&=-\theta_{\rm APS}(h_x,h_y),\nonumber\\
\theta_{\rm APS}(h_x,-h_y)
&=-\theta_{\rm APS}(h_x,h_y)\pmod{2\pi}.
\label{eq:aps_reflections}
\end{align}
We use these identities to complete symmetry-related points of the toron grid, and check them directly at several pairs of points.

We evaluate APS phases on a fixed $4\times4$ toron mesh at radial cutoffs $p=20,24,28$, with $K_{\rm APS}=28$ and 136 radial quadrature points. Table~\ref{tab:aps_sixteen} shows the maximum phase differences over this mesh. Increasing $p$ improves the agreement at $\delta=0.0125$. At the smaller regulator $\delta=0.00625$, an increase of the angular cutoff from $K_{\rm APS}=28$ to $40$ is also needed. These two scans distinguish radial and angular truncation effects.

We also vary one cutoff at a time on four common toron values, as listed in Table~\ref{tab:aps_independent_cutoffs}. The radial convergence is clear at $\delta=0.0125$, while the regulator dependence is nonmonotonic at fixed angular and radial cutoffs. Both tables show the need to increase spectral resolution alongside changes in the heat regulator.

\begin{table}[t]
\begin{ruledtabular}
\begin{tabular}{cccc}
$(p,K_{\rm APS})$ & $\delta=0.025$ & $\delta=0.0125$ & $\delta=0.00625$\\
\hline
$(20,28)$ & $0.01440$ & $0.01552$ & $0.01677$\\
$(24,28)$ & $0.01424$ & $0.00405$ & $0.01467$\\
$(28,28)$ & $0.01419$ & $0.00209$ & $0.01452$\\
$(28,40)$ & --- & --- & $0.00193$
\end{tabular}
\end{ruledtabular}
\caption{Largest absolute APS phase difference from the analytic result over a fixed $4\times4$ toron mesh. All differences are in radians. The last row increases only $K_{\rm APS}$ at $\delta=0.00625$.}
\label{tab:aps_sixteen}
\end{table}

\begin{table}[t]
\begin{ruledtabular}
\begin{tabular}{cccc}
$p$ & $K_{\rm APS}$ & $\delta$ & Max. difference\\
\hline
$16$ & $28$ & $0.01250$ & $0.03465$\\
$20$ & $28$ & $0.01250$ & $0.01097$\\
$24$ & $20$ & $0.01250$ & $0.00631$\\
$24$ & $24$ & $0.01250$ & $0.00237$\\
$24$ & $28$ & $0.01250$ & $0.00237$\\
$28$ & $28$ & $0.02500$ & $0.01063$\\
$28$ & $28$ & $0.01250$ & $0.00093$\\
$28$ & $28$ & $0.00625$ & $0.00619$
\end{tabular}
\end{ruledtabular}
\caption{APS cutoff scans at four common toron values, varying one cutoff at a time. The last column is the largest absolute phase difference from the analytic determinant, in radians.}
\label{tab:aps_independent_cutoffs}
\end{table}

Table~\ref{tab:toronphase} gives representative absolute phases at a single spectral setting, including a toron outside the midpoint grid.

\begin{table*}[t]
\begin{ruledtabular}
\begin{tabular}{cccc}
$(h_x,h_y)$ & Exact & APS & Difference\\
\hline
$(-0.34375,-0.09375)$ & $-1.244132$ & $-1.243257$ & $+0.000876$\\
$(+0.17000,-0.13000)$ & $-0.472273$ & $-0.471563$ & $+0.000709$\\
$(+0.15625,-0.15625)$ & $0$ & $+0.000354$ & $+0.000354$\\
$(+0.28125,-0.21875)$ & $-2.581888$ & $-2.582828$ & $-0.000940$
\end{tabular}
\end{ruledtabular}
\caption{Absolute flat-toron phases from the APS spectrum at $(p,K_{\rm APS},\delta)=(24,28,0.0125)$. All phases and differences are in radians, with differences taken modulo $2\pi$.}
\label{tab:toronphase}
\end{table*}

\subsection{Nonflat phases}

For nonflat fields the operator is represented in a finite space satisfying the APS boundary condition. We diagonalize the coupled boundary Dirac operator and restrict the bulk basis by its APS spectral projector. In the quadratic form we also retain contributions outside the truncated angular window. The radial matrix elements are evaluated with 48 quadrature points, except for a 64-point check in Appendix~\ref{app:gauge}. We compare the exponentiated phase, since an eta invariant can change by an even integer during spectral flow without changing $\exp(-i\pi\eta)$. The Fourier windows used here are the same as for flat fields, with $K_{\rm APS}$ replaced by $L$. For example, at $(L,p)=(6,14)$ the APS-projected quadratic-form matrix has dimension $3888$ for each charged fermion. A dense matrix of this size is diagonalized separately for the flat and nonflat fields, so the numerical cost is considerably higher than for the two-dimensional PV determinant.

For the bulk-extension test, let $r\in[0,1]$ be the disk radius and $n$ the angular Fourier index of a nonzero boundary harmonic. We use the smooth radial profiles
\begin{align}
g^\theta_{n,a}(r)&=[1-s(r)]r^{|n|+2+a}+s(r),\nonumber\\
g^y_{n,a}(r)&=[1-s(r)]r^{|n|+a}+s(r),
\label{eq:bulk_profiles}
\end{align}
for its two gauge-field components, using the same smooth step $s(r)$ as in the metric. The two fillings have $a=0$ and $a=2$. Both profiles are smooth at the disk origin and agree on the boundary. The flat toron contribution is kept fixed.
The radial and angular cutoff dependence of the nonflat relative phase for configuration I is shown in Table~\ref{tab:aps_cutoff}.

\begin{table}[t]
\begin{ruledtabular}
\begin{tabular}{ccccc}
Radial size $p$ & $L=4$ & $L=5$ & $L=6$ & $L=7$\\
\hline
$6$  & $-6.8540$ & $-7.6067$ & $-7.3365$ & ---\\
$8$  & $-2.6482$ & $-3.4031$ & $-3.1271$ & ---\\
$10$ & $-0.8042$ & $-1.5597$ & $-1.2813$ & ---\\
$12$ & $-0.1890$ & $-0.9448$ & $-0.6655$ & $-0.4743$\\
$14$ & $+0.0025$ & $-0.7534$ & $-0.4739$ & $-0.2825$
\end{tabular}
\end{ruledtabular}
\caption{Relative APS phase in milliradians for configuration I at $\max_{i,j}|\phi_{ij}|=0.461$ and $\delta=0.00625$. The analytic value is zero.}
\label{tab:aps_cutoff}
\end{table}

Table~\ref{tab:charge_eta} resolves the relative phase of configuration II by charge at several angular and radial cutoffs. Each entry is the signed contribution to the relative phase in milliradians. The right-handed charge-five contribution has the opposite sign to the left-handed contributions in the sum of eta invariants. The neutral fermion has no relative contribution. The continuum result for the full theory is zero. At these finite cutoffs, the dominant residual comes from the charge-five sector. Its contribution is substantially reduced as $L$ is increased at fixed $p=12$, and again between $(L,p)=(6,14)$ and $(7,14)$. The comparison shows the cutoff dependence of the charge-resolved contributions and their sum. Only the anomaly-free combination is required to have zero continuum relative phase. For the alternative bulk profile at $(L,p)=(6,14)$, the right-handed charge-five contribution is $+0.4395$ mrad, compared with $+0.4437$ mrad in Table~\ref{tab:charge_eta}. The corresponding total phases are $+0.4413$ and $+0.4348$ mrad.

\begin{table}[t]
\begin{ruledtabular}
\begin{tabular}{crrrr}
$(L,p)$ & Left $q=3$ & Left $q=4$ & Right $q=5$ & Total\\
\hline
$(4,12)$ & $-0.0050$ & $-0.0551$ & $+3.6689$ & $+3.6087$\\
$(5,12)$ & $-0.0018$ & $-0.0238$ & $+2.0164$ & $+1.9908$\\
$(6,12)$ & $-0.0015$ & $-0.0080$ & $+0.4783$ & $+0.4689$\\
$(6,14)$ & $-0.0014$ & $-0.0075$ & $+0.4437$ & $+0.4348$\\
$(7,14)$ & $-0.0028$ & $-0.0061$ & $+0.0464$ & $+0.0376$
\end{tabular}
\end{ruledtabular}
\caption{Charge-resolved relative APS phase in milliradians for configuration II with $\max_{i,j}|\phi_{ij}|=0.200$ and $\delta=0.00625$. The same bulk extension is used in every row.}
\label{tab:charge_eta}
\end{table}

\section{Gauge and bulk tests at $Q=0$}
\label{app:gauge}

\subsection{Gauge transformations}

With the compatible real lifts of Sec.~\ref{subsec:gauge_interp}, a gauge transformation of the links induces a continuum gauge transformation of the interpolated connection.  The continuum determinant of the anomaly-free theory is gauge invariant, including its phase. At finite spectral cutoff this property requires care, since a fixed Fourier basis is not preserved by a general gauge transformation.

We first test the determinant magnitude for the unscaled configuration I with toron $(0.17,0.13)$ and $k_{\rm cut}=6$, using $\chi_g(x,y)=0.15\cos[2\pi(x+y)]$. When the same fixed Fourier window is used before and after the transformation, the truncated operator gives a noticeable spurious change. We then represent the transformed operator in a gauge-transformed Fourier basis. The two finite matrices are unitarily equivalent by construction, up to the numerical Fourier representation of the gauge function. At PV mass $M=12$ and $K=3,4,5$, the logarithm of the PV-regularized determinant magnitude changes by $0.512058$, $0.319732$, and $0.210463$ in the fixed basis, but only by $+2.8\times10^{-13}$, $-5.7\times10^{-13}$, and $-3.7\times10^{-12}$ in the transformed basis, which uses the Jacobi--Anger expansion of the gauge transformation. The small residual in the transformed basis checks the implementation. The fixed-basis values show the effect of the truncation.

The APS phase has a more direct test.  We apply $\chi_g(y)=0.04\cos(2\pi y)$ to configuration II with $\max_{i,j}|\phi_{ij}|=0.2$ and toron $(-0.15625,0.09375)$. This gauge transformation extends smoothly into the solid-torus bulk. In the numerical calculation, the bulk connection itself is changed by $A_y(r,\theta,y)\mapsto A_y(r,\theta,y)+\partial_y\chi_g(y)$, with the same pure-gauge term at every radius. We do not reconstruct a new bulk filling from gauge-transformed boundary links. The three-dimensional Dirac operators are assembled separately before and after the transformation in the same fixed Fourier basis. At $(L,p)=(3,6)$ the phase difference is $0.001150$ rad, while at $(4,8)$ it decreases to $0.000408$ rad. The reduction demonstrates the sensitivity of the finite-cutoff APS phase to the spectral resolution. Large gauge transformations require separate tests of the zero-mode conventions and bulk extension.

\subsection{Bulk extensions}

For the anomaly-free $3450$ representation, the continuum APS prescription predicts that the exponentiated phase is independent of the extension of the boundary field into the bulk~\cite{Dai:1994kq,Witten:2019bou}. We test this statement by keeping the boundary connection fixed while changing the radial profile of nonzero Fourier modes in the interior. The two smooth profiles are given in Appendix~\ref{app:aps} and induce the same boundary connection. We calculate the phase separately for each filling. Their finite-cutoff differences measure the stability of the APS spectral calculation. The results at $\delta=0.00625$ are collected in Table~\ref{tab:aps_checks}. For the signed difference we use $\theta_{a=2}-\theta_{a=0}$ modulo $2\pi$.

Table~\ref{tab:aps_checks} shows that the extension difference is sensitive to both $L$ and $p$. For configuration II it becomes particularly small at $(L,p)=(6,14)$, but increases again at $(7,14)$. The relative phases of the anomaly-free combination for both extensions nevertheless move closer to zero as $L$ increases. Agreement between bulk extensions and agreement with the exact phase are therefore distinct tests. Varying $\delta$ between $0.003125$ and $0.0125$ changes these differences less than changing the spectral basis. At $(6,14)$, increasing the radial quadrature from 48 to 64 points changes the extension difference by $2.85\times10^{-7}$ rad.

\begin{table}[t]
\begin{ruledtabular}
\begin{tabular}{lcr}
Test & $(L,p)$ & $|\Delta\theta|$ ($10^{-6}$ rad)\\
\hline
Gauge transformation, II & $(3,6)$ & $1150$\\
Gauge transformation, II & $(4,8)$ & $408$\\
Bulk extension, II & $(4,8)$ & $297$\\
Bulk extension, II & $(4,12)$ & $656$\\
Bulk extension, II & $(5,8)$ & $691$\\
Bulk extension, II & $(5,12)$ & $255$\\
Bulk extension, II & $(6,12)$ & $59$\\
Bulk extension, II & $(6,14)$ & $6.47$\\
Bulk extension, II & $(7,14)$ & $48.0$\\
Bulk extension, I & $(4,12)$ & $336$\\
Bulk extension, I & $(5,12)$ & $258$
\end{tabular}
\end{ruledtabular}
\caption{Finite-cutoff APS gauge and bulk-extension checks. The last column gives the absolute phase change modulo $2\pi$ in units of $10^{-6}$ rad; all tests use $\delta=0.00625$.}
\label{tab:aps_checks}
\end{table}

\section{Numerical methods at $Q=1$}
\label{app:qone}

This appendix collects the spectral and boundary-condition details of the $Q=1$ calculation.

\subsection{Landau basis and PV ratio}

For a charge $q>0$, the $q$ reference zero modes at $Q=1$ can be written as theta functions~\cite{Sachs:1991en}, and each higher Landau level has the same degeneracy. The target zero vectors are obtained numerically from the null space of the rectangular Weyl matrix of Sec.~\ref{subsec:qone_setup}, with their phases and normalizations tracked relative to the reference. Since that matrix has index $q$ by construction, the zero-mode count is not a numerical test of the index theorem.

The overlap matrix in Eq.~(\ref{eq:qone_overlap}) is calculated from the analytic uniform-flux zero modes. It involves neither the numerical null vectors nor the three-dimensional spectrum. For the $x$-dependent profiles it is diagonal in the guiding-center label. Results from quadrature with $1024$ and $8192$ points agree in $\arg\det\Omega^{(q)}$ to better than $10^{-13}$ rad for background O.

For the smooth $x$-dependent backgrounds, we compute the nonzero singular values in each of the $q$ guiding-center blocks, evaluate the primed version of Eq.~(\ref{eq:pv_product}), and form target-to-reference ratios, which are multiplied by the normalized zero-mode Slater determinants at the twelve insertion points. The block calculation agrees with a singular-value decomposition of the full rectangular matrix to better than $6\times10^{-14}$ in tests with 120 Landau levels per charge. For the link-interpolated backgrounds, we retain the couplings between guiding-center sectors induced by the additional Fourier modes.

For reproducibility, each insertion set is generated from an integer $s$. For each charge $q$ we use the PCG64 pseudorandom generator with integer seed $s+q$. The first $q$ independent uniform draws in $[0,1)$ are the $x$ coordinates, and the next $q$ are the $y$ coordinates. Sets 1--4 in Table~\ref{tab:qone_positions} use $s=71,89,137,223$.

\subsection{Cylinder and zero-mode gluing}

The cylinder coordinate $\tau\in[0,1]$ runs from the uniform-flux reference to the target, with the interpolation of Eq.~(\ref{eq:qone_bulk_interp}). Other orderings of the transverse and pure-gauge deformations are used for the bulk-path test below. In $H_Y=i\Gamma^\tau\partial_\tau+B(\tau)$, $\Gamma^\tau$ is the Clifford matrix normal to the boundary. The cylinder spinor is expanded in $p$ normalized Legendre polynomials. With the rectangular Landau-basis matrix $D(\tau)$ of $\mathcal D_{L,q}$, of size $qn_{\rm max}\times q(n_{\rm max}+1)$, where $n_{\rm max}=\lfloor E_\eta/q\rfloor$, we use
\begin{equation*}
\Gamma^\tau=\begin{pmatrix}1&0\\0&-1\end{pmatrix},\qquad
B(\tau)=\begin{pmatrix}0&D(\tau)^\dagger\\D(\tau)&0\end{pmatrix},
\end{equation*}
where the diagonal entries of $\Gamma^\tau$ denote identity matrices on the two Landau subspaces. The tangential operator is
\begin{equation*}
\mathcal B=-i\Gamma^\tau B=\begin{pmatrix}0&-iD^\dagger\\iD&0\end{pmatrix}.
\end{equation*}
Let $v_j$ and $u_j$ be the right and left singular vectors of $D$ with nonzero singular values $\sigma_j$. The corresponding eigenvectors of $\mathcal B$ are $(v_j,\pm i u_j)^T/\sqrt{2}$, with eigenvalues $\pm\sigma_j$. We retain the negative eigenspace at $\tau=0$ and the positive eigenspace at $\tau=1$. We rotate the target frame $Z_1$ by the unitary polar factor of $Z_1^\dagger Z_0$, making the aligned overlap positive Hermitian. In these frames our reference calculation takes $V=e^{0.37\pi i}\mathbf 1_q$. The unitary-variation checks change this map while keeping the boundary fields fixed.

The orientation of the gluing map in Eq.~(\ref{eq:qone_zero_gluing}) can be checked in a one-dimensional example. For a single zero-mode channel with $H_Y=i\partial_\tau$ and $\Psi(1)=e^{i\theta}\Psi(0)$, the eigenvalues are $2\pi n-\theta$. For $0<\theta<2\pi$,
\begin{equation}
\eta(0)=\frac{\theta}{\pi}-1,\qquad
e^{-i\pi\eta}=-e^{-i\theta}.
\label{eq:qone_1d_eta}
\end{equation}
The constrained polynomial calculation reproduces this phase, which fixes the orientation needed when the eta phase is multiplied by the Slater determinants.

Numerically we impose the boundary constraints, construct a basis of their null space, and diagonalize the positive quadratic form $\langle H_Y\Psi,H_Y\Psi\rangle$ in it, as in the zero-flux calculation [Eqs.~(\ref{eq:eta_estimator_main}) and (\ref{eq:projection_diff})]. The relative eta is obtained by subtracting the uniform reference value at common cutoffs. For the large link-interpolated matrices we use an orthonormal polynomial basis adapted to the endpoint constraints and split the matrices into invariant guiding-center and inversion-symmetry blocks where applicable. This leaves the finite Galerkin space unchanged.

\subsection{Gauge, unitary, and bulk checks}

Table~\ref{tab:qone_u_check} lists the residual change in the combined eta and zero-mode phase under two changes of the gluing map, at $E_\eta=120$ and $p=24$. In these tests the endpoint frames and the reference normalization are kept fixed while the gluing map is varied. For $\delta=0.003$ and $0.006$, both residuals are below $2\times10^{-7}$ rad in absolute value.

\begin{table}[t]
\begin{ruledtabular}
\begin{tabular}{lcc}
$\delta$ & Uniform rotation & Det.-preserving\\
\hline
$0.00075$ & $-9.05\times10^{-6}$ & $-4.57\times10^{-6}$\\
$0.0015$  & $+5.63\times10^{-7}$ & $+5.11\times10^{-7}$\\
$0.003$   & $-4.89\times10^{-8}$ & $+1.89\times10^{-7}$\\
$0.006$   & $-7.03\times10^{-8}$ & $+1.78\times10^{-7}$
\end{tabular}
\end{ruledtabular}
\caption{Residual change in the combined eta and zero-mode determinant phase (radians) under two changes of $V$, at $E_\eta=120$ and $p=24$. The uniform rotation changes all gluing eigenphases by $0.23$ rad; the second rotation preserves $\det V$.}
\label{tab:qone_u_check}
\end{table}

For the bulk test in background B, we compare the interpolation in which the first harmonic of $\varphi$ is turned on before the second harmonic and $\chi$, with the reverse ordering. We use $p=24$ and $\delta=0.003$, and compare $E_\eta=120$ and $180$ in Table~\ref{tab:qone_bulk_charge}.

Joining the two paths with opposite orientations gives a closed manifold $Y_{\rm cl}$. By the APS index theorem~\cite{Atiyah:1975jf,Alvarez-Gaume:1984zst}, the eta-phase difference is determined by the relative Chern--Simons integral
\begin{equation*}
I_{\rm CS}^{\rm rel}=\int_{Y_{\rm cl}}(2\alpha\wedge F^{\rm ref}+\alpha\wedge d\alpha),
\end{equation*}
where $\alpha=A-A^{\rm ref}$ is globally defined even at nonzero flux. The first term vanishes since $\alpha_\tau=0$ and $F^{\rm ref}=2\pi\,dx\wedge dy$. Here
\begin{equation}
A_x^{(c)}=-2\pi c\sin(2\pi x),\qquad
A_y^{(a)}=2\pi a\sin(2\pi x).
\end{equation}
The first path turns on $A_y^{(a)}$ before $A_x^{(c)}$, and the second reverses this order. Thus
\begin{align}
I_{\rm CS}^{\rm rel}&=\int_{Y_{\rm cl}}\alpha\wedge d\alpha
=2\int_{T^2}d^2x\,A_x^{(c)}A_y^{(a)}\nonumber\\
&=-4\pi^2ac.
\label{eq:qone_cs_difference}
\end{align}
The second harmonic, proportional to $b$, integrates to zero by Fourier orthogonality. With our boundary orientation the phase convention in Eq.~(\ref{eq:qone_numerical_phase}) gives $\Delta\Theta_q=-(q^2/4\pi)I_{\rm CS}^{\rm rel}=q^2\pi ac$.

The $Q=1$ heat-kernel regulator has the same definition as in Eq.~(\ref{eq:eta_estimator_main}), with dimensionless $\delta\mu_j$. Since the torus side length is set to one, we quote $\delta$ in the corresponding squared-length units. Tables~\ref{tab:qone_eta} and \ref{tab:qone_u_check} show the dependence of the phase and gluing checks on $\delta$ at finite basis size. A simultaneous study of $E_\eta$, $p$, and $\delta$ will be needed to determine the spectral limit.

Finally we apply opposite pure-gauge deformations $\chi(x)=\pm0.025\cos(2\pi x)$ to the same background B. The exact phases are $\pm0.263915477$ rad, while the numerically assembled phases at $E=600$, $M=24$, $E_\eta=120$, $p=18$, and $\delta=0.00075$ are $+0.263914432$ and $-0.263917063$ rad. Their difference agrees with the analytic gauge-phase change within $5.5\times10^{-7}$ rad. The magnitudes also agree under this sign reversal, as expected for a cutoff artifact that is even in $\chi$ (Sec.~\ref{sec:qone}).

\bibliographystyle{utphys}
\bibliography{ref}

\end{document}